\documentclass[nofootinbib,superscriptaddress,a4paper,twocolumn, floatfix]{revtex4-2}
\usepackage[english]{babel}
\usepackage[utf8]{inputenc}

\usepackage{amsthm}
\usepackage{amssymb}
\usepackage{mathtools}
\usepackage[left=23mm,right=13mm,top=35mm,columnsep=15pt]{geometry} 

\usepackage{subfigure}
\usepackage{siunitx}
\usepackage[dvipsnames]{xcolor}
\usepackage[ruled,vlined]{algorithm2e}
\usepackage{amsmath, amssymb} % For math symbols (\mathbb, \mathbb{R}, norms, etc.)
\usepackage{makecell}

\usepackage{tikz}
\usetikzlibrary{calc}
\usetikzlibrary{automata,positioning}
\usetikzlibrary{arrows.meta, positioning, shapes, fit}

\usepackage[pdftex, pdftitle={Article}, pdfauthor={Author}]{hyperref}
\hypersetup{
    colorlinks,
    linkcolor={red!50!black},
    citecolor={blue!50!black},
    urlcolor={blue!80!black}
}

\usepackage{booktabs}
\usepackage{multirow}
\usepackage{listings}

\newcommand{\expect}[1]{\langle #1 \rangle}

\begin{document}

\title{A Transferable Machine Learning Approach to Predict Quantum Circuit Parameters for Electronic Structure Problems} 

% alphabetic order when equal contribution
\author{Davide Bincoletto}
\thanks{authors contributed equally}
\affiliation{{Institute for Computer Science, University of Augsburg, Germany }}

\author{Korbinian Stein}
\thanks{authors contributed equally}
\affiliation{{Institute for Computer Science, University of Augsburg, Germany }}

\author{Jonas Motyl}
\affiliation{{Institute for Computer Science, University of Augsburg, Germany }}

\author{Jakob~S.~Kottmann}
\email[E-mail:]{jakob.kottmann@uni-a.de}
\affiliation{{Institute for Computer Science, University of Augsburg, Germany }}
\affiliation{{Center for Advanced Analytics and Predictive Sciences, University of Augsburg, Germany }}

\date{\today} % Leave empty to omit a date

\begin{abstract}
The individual optimization of quantum circuit parameters is currently one of the main practical bottlenecks in variational quantum eigensolvers for electronic systems. 
To this end, several machine learning approaches have been proposed to mitigate the problem. However, such method predominantly aims at training and predicting parameters tailored to individual molecules: either a specific structure, or several structures of the same molecule with varying bond lengths.
This work explores machine learning based modeling strategies to include transferability between different molecules.
We use a well investigated quantum circuit design and apply it to model properties of hydrogenic systems where we show parameter prediction that is systematically transferable to instances significantly larger than the training instances. 
\end{abstract}

\maketitle

\section{Introduction}

The electronic structure problem for molecular systems is a central problem in quantum chemistry, 
providing important insights for real-world applications like materials science and drug discovery.
While classical computation methods are sufficiently fast for smaller systems, 
the field of quantum computing promises to efficiently find the resulting ground state energy of larger quantum chemical systems.~\cite{quantum-comp-chem}

The problem can be briefly stated as finding the lowest eigenvalue of a given second-quantized Hamiltonian -- a formally $2^{2N}\times 2^{2N}$ matrix given in a sparse format requiring only $O(N^4)$ numbers. Here $N$ denotes the number of Fermionic sites, or orbitals. Since electrons have a spin, each orbital can be occupied by a spin-up or spin-down electron (or both). A natural mapping to qubit systems is then to use $2N$ qubits to represent the occupation of the $N$ orbitals by spin-up and spin-down electrons.
Quantum Algorithms, like the quantum phase estimation~\cite{aspuru2005simulated}, project a given input state onto an eigenstate of the electronic Hamiltonian, where the success probability is proportional to the input state and the corresponding eigenstate. 
This is however an expensive protocol~\cite{reiher2017elucidating, vonburg2021quantum}, where recent improvements bring the possibility of successful future applications closer to reality.~\cite{rocca2024reducing, cortes2024assessing, ollitrault2025improving}
It is therefore advantageous to have methods that can prepare adequate initial states that approximate electronic ground states.
Other algorithms, like prolate diagonalization~\cite{stroschein2025groundexcitedstateenergiesanalytic}, imaginary-time evolution~\cite{motta2020determining} or efforts to simulate quantum dynamics~\cite{Langkabel2022} also profit from efficient initial state preparations.\\

A class of algorithms capable of direct state preparation are Variational Quantum Eigensolvers (VQE)\cite{peruzzo2014variational} that use the variational principle to optimize parameters of specific circuit designs. \cite{anand2022quantum, tillyVariationalQuantumEigensolver2022, cerezo2021variational, bharti2022noisy}
While there are good arguments, that VQEs alone will not reach verifiable quantum advantage they are still useful for the task of initial state preparation -- once the circuit is trained, the preparation cost is typically cheap.
In a nutshell, a VQE protocol takes a parametrized quantum circuit and couples it to a classical optimizer to find the parameters that minimize the expectation value with respect to an electronic Hamiltonian. We provide further technical details in Appendix \autoref{c:VQE}. Over the years, several circuit designs have emerged~\cite{anand2022quantum, romero2018strategies, lee2018generalized, kottmann2022optimized, kottmannMolecular2023, burton2022exact, burton2024accurate}, most are however not developed enough to be applicable in a black-box fashion. One candidate with such property is the separable pair approximation (SPA)~\cite{kottmann2022optimized}, which has proven to be a robust method in several applications~\cite{weber2022toward, schleich2021improving, gil2025sharc, santosHybrid2024, schleich2023partitioning, kottmann2024quantum} and can be extended through various strategies~\cite{lee2018generalized, kottmannMolecular2023, burton2024accurate}. This will be employed throughout the work. Further details are presented in Appendix \autoref{c:SPA}.

The classical optimization step can be a notoriously difficult task~\cite{mcclean2018barren, bittel2021training} for unstructured approaches. This has led to multiple attempts at using alternative techniques, e.g., machine learning, to model VQE parameters with the goal of improving reliability in finding optimal solutions.
Modeling quantum circuit parameters has shown success in various related works.\textcite{zhang2020collective} started with collective optimization of groups of Hamiltonians originating from the same molecule at various conformations. Later \textcite{MetaVQE} introduced the Meta-VQE, addressing the problem on a more general level: minimizing the parameters of a given parametrized Hamiltonian. This approach is inspired by quantum machine learning (QML) algorithms and extends beyond quantum chemistry applications. Alternatively, several works~\cite{related-ceroni2023generatingapproximategroundstates, related-warm-starting-vqe} employed classical neural network or machine-learning related approaches to directly predict circuit parameters or whole circuits~\cite{related-generative-quantum-eigensolver} from the Hamiltonian.
More recently, \textcite{bensoussan2025accelerq} introduced a framework that applies the XGBoost ML predictor model for optimizing the hyperparameters of a broad class of Quantum Eigensolvers.

Despite numerous investigations, we are currently not aware of a method that accurately leverages molecular structure and is capable of generalizing to large-size systems. This work explores multiple approaches to model variational quantum parameters of system-adapted quantum circuits based on atomic coordinates. We extend upon the basis of the Meta-VQE concept with classical machine learning techniques to predict parameters that enable initialization of the ansatz circuit in an optimal state. We evaluate our models on hydrogenic systems up to H$_{12}$ (at this time, limited due to simulation cost), with a major focus on to demonstrate transferability. In fact, while most of the methods require training for different sizes of parametrized Hamiltonians, our method aims at building a scalable model that can extend to multiple molecule sizes. We refer to~\cite{kottmannMolecular2023, kottmann2024quantum} for potential transferability between hydrogenic model chemistries and more general classes of molecules. 

This article is structured as follows. In \autoref{c:main}, we introduce the data generation and the modeling approaches. In \autoref{s:evaluation} we show the results of our experimentation for multiple datasets. Finally, we discuss our findings and list several topics for future work in \autoref{c:conclusion}.

\section{Methodology}
\label{c:main}

To tackle the parameter prediction problem, we developed three architectures. For the first model we trained a Graph Attention Network (GAT) \cite{gat}
with the aim of using the graph structure of the molecule as a basis for a dynamic prediction model, leveraging the message passing between adjacent nodes.
Previous applications to chemistry have shown promising results on multiple molecular properties at the graph level, including but not limited to the ground state energy. \cite{related-neural_message_passing,related-schnet,related-psiformer-attention}
Our algorithms employ a similar problem structure, representing molecular geometries as graphs, where atoms correspond to vertices and the bonds to edges.

The second and third models are based on the Schrödinger's Network (SchNet)~\cite{schnet} -- a neural architecture specifically designed for learning molecular representations. It employs continuous filter convolutions, which enables the accurate modeling of interactions at small distances. The native model predicts ground state energy for a mixed set of molecules. Since the problem we are focusing on is similar, we can use the SchNet as backbone and employ the core structure of the architecture for modeling variational parameters, which then result in the ground state energy. With a new head for the backbone we are able to train two different models based on the SchNet architecture with slightly different architecture approaches, which have different weighting on the employed graphs of our ansatz. The models have been developed sequentially in the order presented with the goal to further refine their capabilities and allow training with reduced training sets. An overview of all the models features is showcased in Table \ref{tab:models-overview}.

\begin{table*}[ht]
  \caption{Overview of the examined neural network architectures. GAT and Linear SchNet models use identical training data and preprocessing, while the Mixed SchNet model employs a smaller, mixed dataset with coordinate reordering based on graph matching.}
  \label{tab:models-overview}
  \centering
  \resizebox{0.8\textwidth}{!}{
      \begin{tabular}{c|c|c|c}
        \toprule
        Model & Training dataset & \# Parameters & Input Preprocessing  \\
        \hline
       Graph Attention Network (GAT)  &  230k linear H$_4$ &   302,625  & \makecell{euclidian distance matrix\\ with angles} \\ \hline
       Linear SchNet   &  230k linear H$_4$ &   28,273  & \makecell{euclidian distance matrix\\ with angles} \\ \hline
       Mixed SchNet   &  \makecell{1.000 linear H$_4$  \\ 2.000 random H$_6$ } &   472,450  & \makecell{reorder coordinates according\\ to perfect matching graph} \\
        \bottomrule
      \end{tabular}
    }
\end{table*}

\subsection{Data generation}
\label{s:method}
\begin{figure*}[ht]
  \begin{center}
    \includegraphics[width=0.7\textwidth]{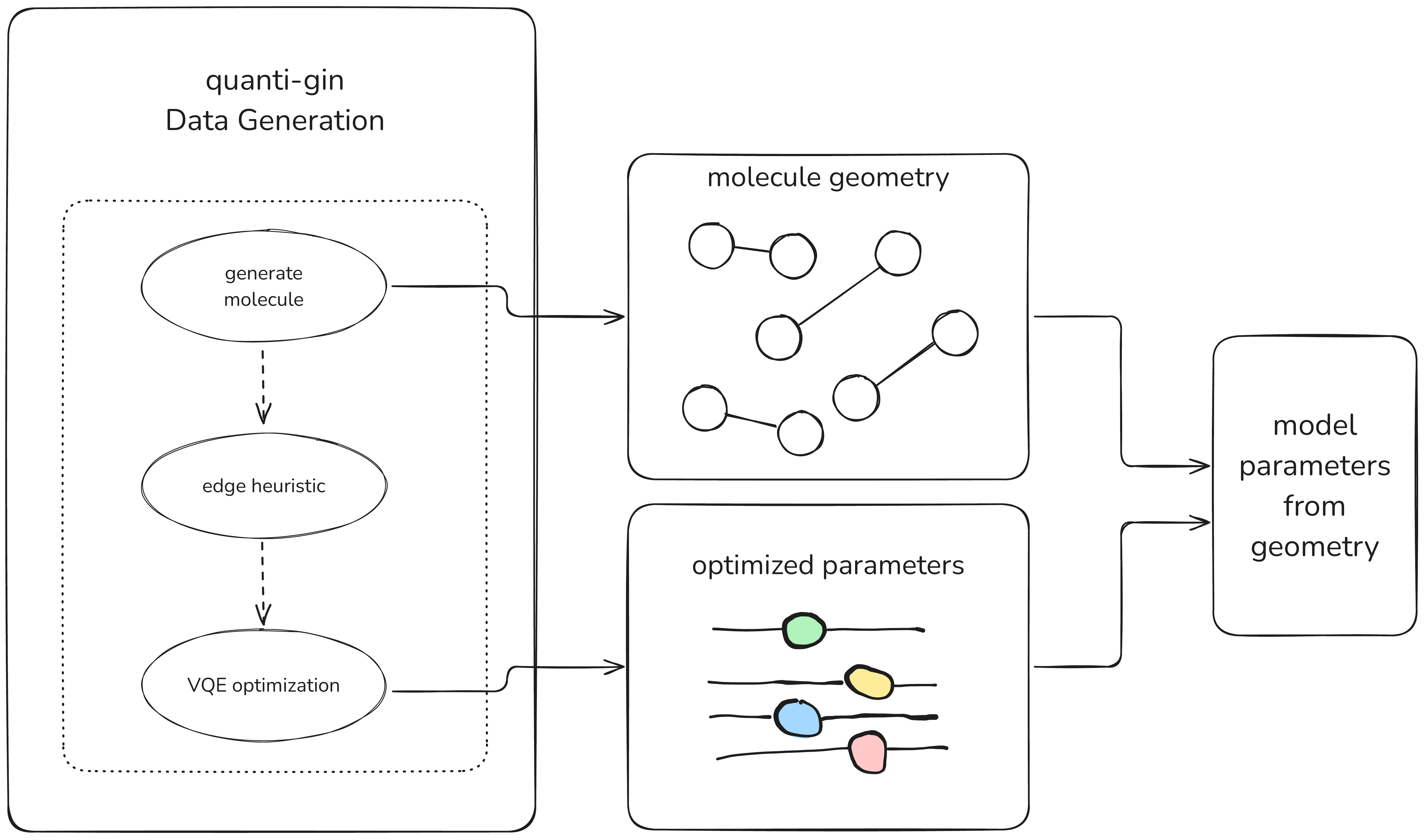}
  \end{center}
  \caption{Parameter optimization flow - generating randomized molecule geometries and their optimized parameters, which are the input into the model.}\label{fig:overview-drawing}
\end{figure*}

For parameter modeling and analysis, we made use of $\textsc{quanti-gin}$\cite{steinNylser2025} to generate datasets of molecular geometries containing Hamiltonians, quantum circuits in the form of the separable pair ansatz (SPA)~\cite{kottmann2022optimized} and the corresponding optimized parameters following the strategies of Ref.~\cite{kottmannMolecular2023}. The workflow is illustrated in \autoref{fig:overview-drawing}.
The procedure for generating a datapoint with index $i$ (out of $n$ total) is as follows:
\begin{enumerate}
    \item Generate the coordinates $C$ according to the molecular structure following Algorithm \ref{alg:random-coordinates}. See Appendix for alternative structures.
    \item Estimate the optimal chemical graph (Lewis formula) for the coordinates $C$. We used scaled Euclidean distances as edge weights and heuristically determined a perfect matching graph $G$ with minimal edge weight.
    \item Construct the circuit ansatz $U_{\text{SPA}}$ employing the graph $G$ and compute the corresponding orbital-optimized Hamiltonian $H_{\text{opt}}$.
    \item Minimize the expectation value $\langle U_{\text{SPA}} | H_{\text{opt}} | U_{\text{SPA}} \rangle$ with a VQE, to obtain the corresponding energy $E_{\text{SPA}}$ and parameters $\theta$.
    \item Normalize the angles and store the set $(C, H, G, E_{\text{SPA}}, \theta)$ as one instance of the dataset.
\end{enumerate}

Here we applied a constraint on the coordinates $C$ generated by the randomized procedure (Algorithm~\ref{alg:random-coordinates}). Starting from the origin, each new atom is placed at a random point with a distance between $0.5$ and $2.5$~\si{\angstrom} from the previous existing atom, ensuring a minimum separation of $0.5$~\si{\angstrom}. This avoids clustering of very small inter-atomic distances as well as generating dissociated structures with no relevant physical interactions. See the code in the Appendix for a step-by-step example using the \textsc{tequila}~\cite{tequila} library.

In the experimentation we generated two training datasets: a large one, made of 230.000 linear H$_4$ instances, and a small one, made of 1.000 linear instances of H$_4$ and 2.000 random instances of H$_6$. The purpose of the large dataset is mainly to get a good estimate of the best possible result when only training on linear H$_4$ instances. Where the smaller training set was chosen to demonstrate that comparable, or improved, quality can be obtained with significantly smaller training sets that still contain only small instances. For the evaluation we produced several datasets of 500 random instances with varying molecule sizes (ranging from H$_2$ to H$_{12}$). Here we chose a minimal basis approach (one orbital, or two qubits, for each hydrogen atom) and refer to other works~\cite{kottmann2020reducing, schleich2021improving,Langkabel2022} regarding improved problem encodings. 
In this work, we focus on hydrogenic systems because they are relatively simple to treat, while still containing fundamental geometrical structures that can be leveraged in the study of non-hydrogenic systems, as shown in previous works. \cite{kottmannMolecular2023, BincolettoKottmann2025}

\begin{algorithm}[H]
\caption{Random Coordinates Generation with Minimum Separation}
\label{alg:random-coordinates}
\KwIn{
    Number of atoms $n \in \mathbb{N}$, \\
    maximum displacement $d_{\max} \in \mathbb{R}^+$, \\
}
\KwOut{
    Coordinate matrix $C \in \mathbb{R}^{n \times 3}$
}
\BlankLine
Set $C \gets \{(0,0,0)\}$\; 

\For{$i \gets 1$ \KwTo $n-1$}{
    Select a reference point $\mathbf{r}_{\mathrm{prev}} \in C$ uniformly at random\;
    $\delta_{\min} \gets 0$\;
    
    \While{$\delta_{\min} \leq 0.5$}{
        Sample $\mathbf{u} \sim \mathcal{U}([0,1]^3)$ at random\;

        $\mathbf{r}_{\mathrm{new}} \gets \mathbf{r}_{\mathrm{prev}} + d_{\max} \cdot \mathbf{u} $

        $\delta_{\min} \gets \min_{\mathbf{r} \in C} \|\mathbf{r}_{\mathrm{new}} - \mathbf{r}\|_2 $
    }
    Append $\mathbf{r}_{\mathrm{new}}$ to $C$\;
}
\Return $C$\;
\end{algorithm}

\subsection{Modeling approaches}
\label{s:implementation}

Our goal is to build a model which, given the features derived from the molecule (positions and atom types), outputs a value for the parameters (angles) of the VQE circuit.
As these angles are continuous values, $\theta \in [-2\pi, 2\pi]$, this is a regression problem.
We chose \textsc{PyTorch}~\cite{pytorch} as deep learning framework, due to its wide use in the field.
It provides a large amount of built-in primitives, methods and utilities that can be used in the construction of neural networks.
This was combined with \textsc{PyTorch Geometric}~\cite{pytorch-geometric} for additional functionalities.

\begin{figure}
    \centering
    \subfigure[]{\includegraphics[width=0.4\textwidth]{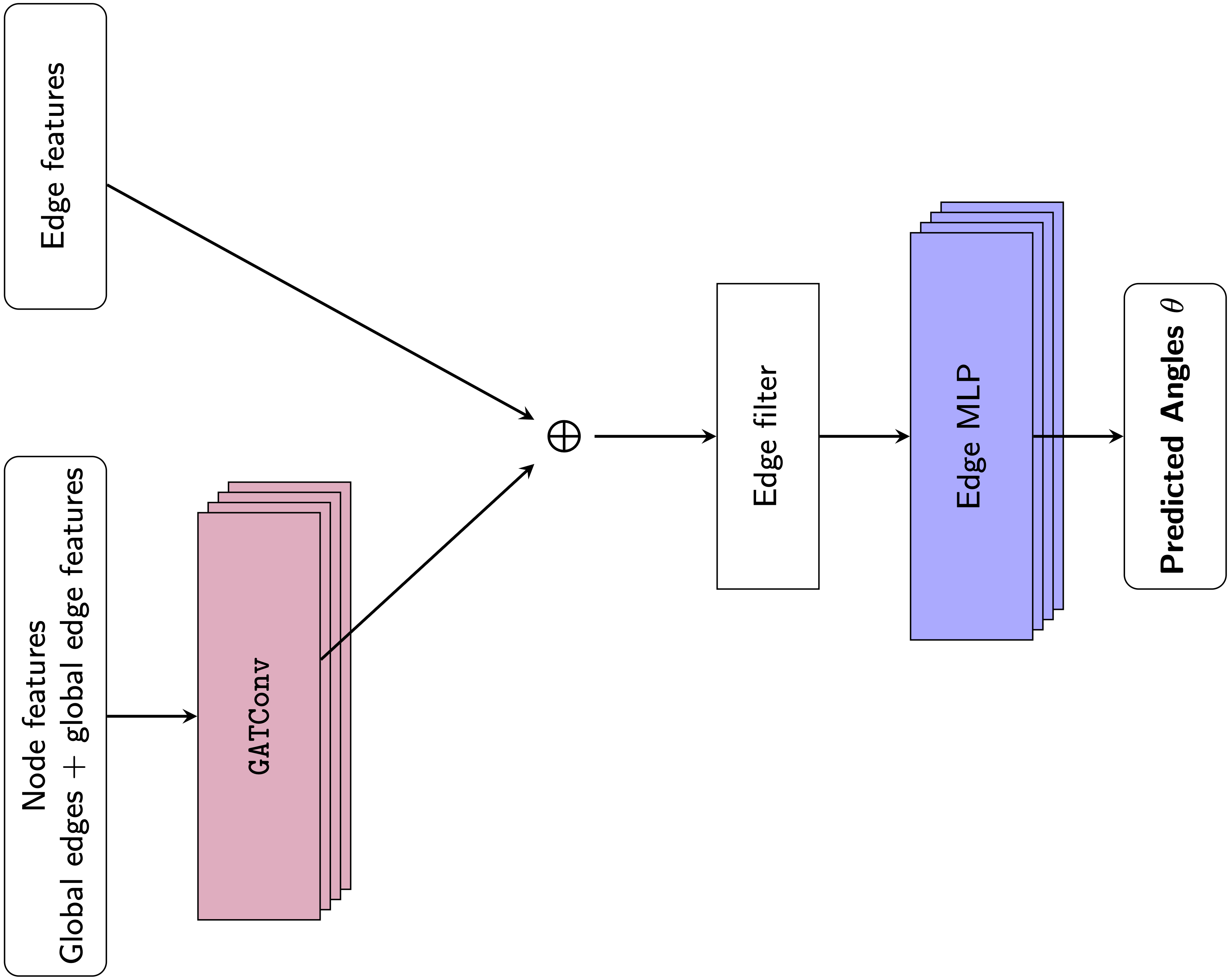}\label{fig:gat-gnn-layout}}
    \subfigure[]{\includegraphics[width=0.4\textwidth]{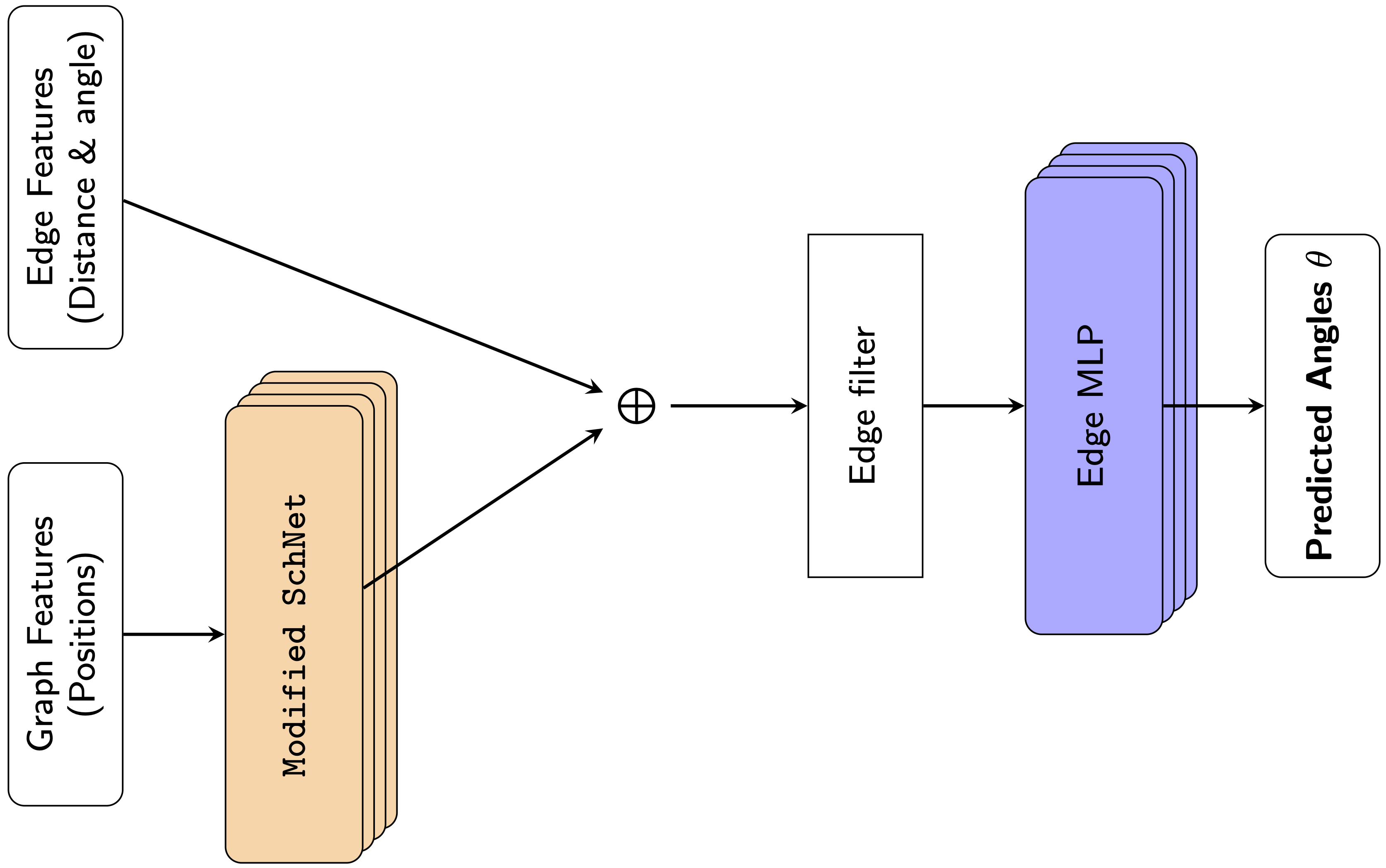}\label{fig:schnet-layout}}
    \subfigure[]{\includegraphics[width=0.36\textwidth]{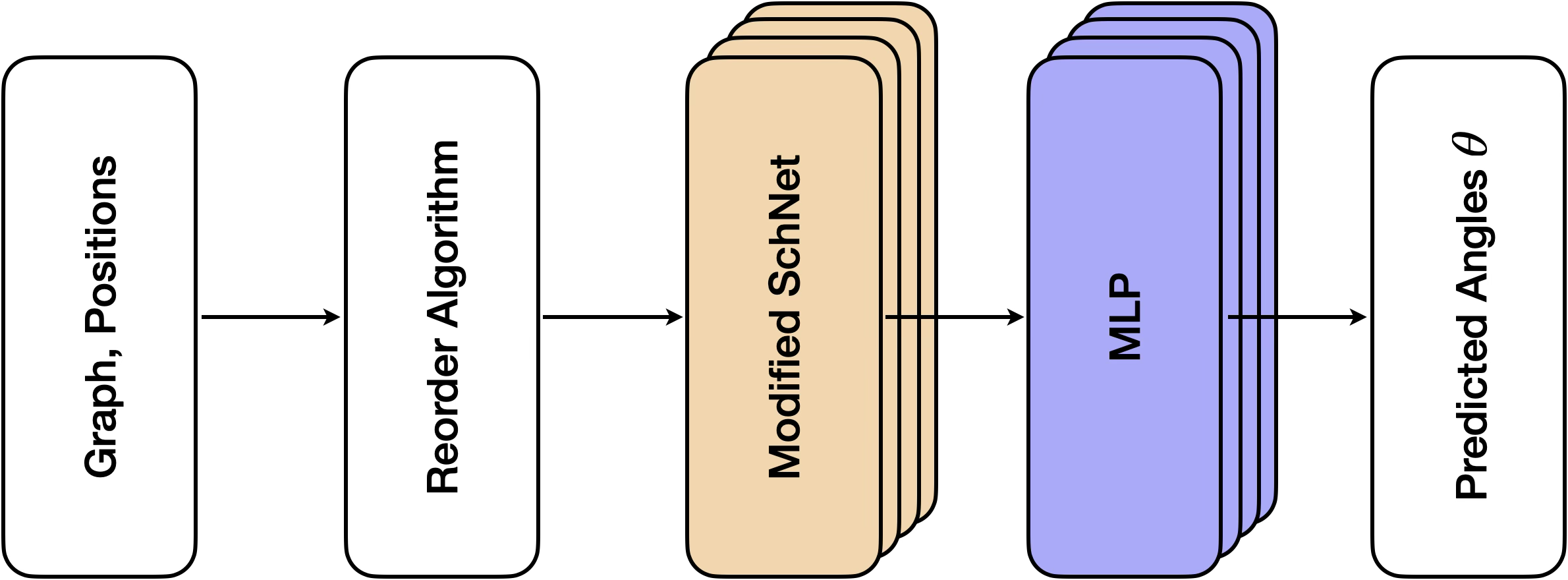}\label{fig:mixed-schnet-layout}}
    \caption{Network diagram of the examined architectures for angle prediction. (a) GAT architecture that processes node and edge features through GatConv layer. (b) Linear SchNet-based model that explicitly incorporates geometric features including atomic positions, interatomic distances, and angles. (c) Mixed SchNet-based model featuring position reordering. All three architectures generate predicted angles $\theta$ through multi-layer perceptrons (MLPs).}
\end{figure}

\subsubsection{Graph Attention Network} \label{sec:gatconv}

Due to the recent success of attention mechanisms ~\cite{related-attention-is-all-you-need}, we decided to use this approach as the foundation for our GAT model.
Such mechanisms have demonstrated effectiveness in capturing complex dependencies between elements in a sequence or graph, thus, we assume they are suitable for modeling molecular interactions.~\cite{survey-attention-models-graph}
\textsc{PyTorch Geometric} provides a Graph Attention Convolution layer (GATConv) implementation, which applies the attention mechanism on node features using message passing to their respective edges.~\cite{gatconv}
Here, message passing refers to the process where each node updates its representation by aggregating and transforming feature information (messages) from its neighbors according to the graph structure.
This approach allows the model to assign different importance weights to neighboring atoms based on their features and relative positions, 
which is crucial for accurately capturing the interactions that determine optimal circuit parameters.
As we use the layer like a black box for the implementation, we will only describe the architecture of the model built around the graph attention convolution.
For more implementation details on the graph attention implementation, refer to~\cite{gatconv}.
We provide a visual overview of the graph-attention GAT in Figure \autoref{fig:gat-gnn-layout}.

Through the combination of the atom coordinates and edges by our heuristic, we already have a graph that we can apply to the GAT, but there are some important factors to consider.
For one, the given graph only contains paired nodes, which, given the edge-based message passing in the GATConv layer,
would lead to only the paired atoms exchanging information.
Additionally, using 3-dimensional coordinates as node features to learn the relationships introduces a feature of high dimensionality. When limited to them, this is likely to greatly increase the required amount of data, therefore also the time and resources required to train the model. For this reason, we perform the following pre-processing and enhancement steps on the molecule graph data before passing it into the model:
\begin{enumerate}
  \item Generate global edges with the same edge heuristic from the data generation routine.
  \item Generate global and non-global edge features, for edges $\vec{e}$, with euclidean distance and angle: 
    \[
    \vec{e}_{uv} = \left( \| \mathbf{x}_u - \mathbf{x}_v \|_2,\ \theta_{uv} \right)
    \]
    where $\mathbf{x}_u, \mathbf{x}_v$ are the positions of nodes $u$ and $v$, and $\theta_{uv}$ is the angle of the corresponding edge.
  \item For each feature set $\vec{v} = (\mathbf{x}_u, \mathbf{x}_v, \vec{e}_{uv})$ perform normalization of the respective components:
    \[
      \vec{v} = \frac{\vec{v} - min(\vec{v})}{max(\vec{v})}
    \]
\end{enumerate}
Importantly, we keep the node features as the bare coordinates of the atoms, as to leave a possibility for the model to extract useful information from coordinates and their correlations.

We first pass the node features, global edges and their generated features through one or more GATConv layers, whose output $x$ is an internal representation of node features,
modeling interactions between the atoms that are in close proximity.
Next, we perform the combination of edge features and node representation $x$, called the edge filter, which uses the non-global edges $uv$ from the edge heuristic described above.
In this step, for each edge $uv$, we concatenate the node features $x_u$ and $x_v$ and the corresponding edge features $\vec{e}_{uv}$:
\[
  \vec{c}_{uv} = \vec{x}_u \oplus \vec{x}_v \oplus \vec{e}_{uv}
\]
Finally, we pass the concatenated features through a Multi-Layer Perceptron (edge MLP) with two linear layers, 
scaling up from the feature dimension of $\vec{c}_{uv}$ to $256$, a ReLU activation function, 
and a final linear layer that scales down to dimension $1$.
As the concatenated edge features $\vec{c}_{uv}$ have dimension $[E, N]$, where $E$ is the count of edges and $N$ the feature dimension, 
we obtain the predicted angles for the circuit $\vec{\theta} \in \mathbb{R}^E$ for every edge between determined pairs.

\subsubsection{Schrödinger Network Backbone}\label{s:schnet-gnn}
While the architecture described in the previous section has the capability to model the impact of inter-pair influences, it is not tuned specifically for predictions of molecular graph representations. For our second architecture, we turned to SchNet. \cite{schnet}
This approach is based on modeling the aforementioned interactions using continuous filter convolutions, instead of discretized grids that the distances need to be fitted to, which allows for the modeling of small distance impacts.
Based on the successful application of SchNet, in order to model quantum chemical properties on molecules with a similar input structure, we propose a new architecture using a modified SchNet as a backbone.
This architecture shares the pre-processing and the final layer with the previously described GAT, but replaces the GATConv layers for modeling interatomic interactions. The network diagram is shown in Figure \autoref{fig:schnet-layout}.

In the original SchNet, the forward pass takes the coordinates $C$ and applies a distance expansion using radial-basis functions. These extend the dimensionality of the distance feature from 1 to 50, passing the distance through 50 Gaussian functions with different centers. After that, the embedded information gets passed to a series of three residual blocks, inspired by ResNet \cite{resnet}. At each layer, our molecule is represented in an atom-wise manner, analogous to pixels in an image. The key mechanism enabling this representation is a series of interaction blocks, which iteratively update atomic feature vectors based on the embedded molecular geometry.

Because we want to process the node representations directly, for concatenating them with the edge information features, 
we cut off the original SchNet after the interaction blocks and add a custom head to it.
We combine the internal node representations with the edge features, where the latter is first passed through the distance expansion layer:
\[
  \vec{c}_{uv} = \vec{x}_u \oplus \vec{x}_v \oplus {f}(\vec{e}_{uv})
\] 
here, $\vec{x}$ are the SchNet node representations and $f(\vec{e})$ the edge features passed through the distance expansion, based on the edge heuristic.
After this combination, we pass the features through a simple edge MLP, again reducing down the dimensionality of the output to match the edges and parameters count in the final VQE circuit.

\subsubsection{Mixed Schrödinger Network}\label{s:mixed-schnet}
Based on the experience with the two previous models, we constructed a third model trained on a significantly smaller dataset including mixed instances (linear and randomized instances of H$_4$ and H$_6$).
For this we used SchNet as a backbone but applied a different head. After the residual interactions, in fact, we take the embedded representation and build new layers on top, see \autoref{fig:mixed-schnet-layout}.
As input we have the coordinates $C = \{ \mathbf{r}_1, \mathbf{r}_2, \dots, \mathbf{r}_n \} \subset \mathbb{R}^3$ and the corresponding perfect matching graph $G$. 
If we want to predict the angles for our SPA circuit we follow the steps:
\begin{enumerate}
    \item Reorder $C$ to according to the perfect matching graph $G$, e.g., $C = \{ \mathbf{r}_1, \mathbf{r}_2, \mathbf{r}_3, \mathbf{r}_4 \} \rightarrow \Tilde{C} = \{ \mathbf{r}_1, \mathbf{r}_4, \mathbf{r}_2, \mathbf{r}_3 \}$ for the graph of H$_4$.

    \item Process the coordinates $\Tilde{C}$ to obtain pairwise embedding with SchNet backbone. This is done by increasing the dimensionality of the available information for each pair to a parameterized hidden dimension, e.g., $s(\mathbf{r}_1, \mathbf{r}_4) \rightarrow X^{1,4}$.

    \item Shape the output of the backbone to the amount of angles we need with a MLP. The circuit ansatz needs one angle for each electron pair, e.g., $\text{MLP(}X^{1,4}) \rightarrow \theta^{1,4}$
\end{enumerate}
Here, the processing steps in the forward pass give the perfect matching graph more emphasis on the output of the model compared to previous architectures.
After concatenating the angles of all layers into one tensor, we have a suitable representation for training and evaluating our model. The network diagram is shown in Figure \autoref{fig:mixed-schnet-layout}.

This architecture differs slightly from the previously mentioned linear SchNet model and notably was trained on mixed (linear and random) instances. The training dataset for this model contains 1.000 linear H$_4$ instances and 2.000 random H$_6$ instances.

\section{Evaluation}
\label{s:evaluation}
There are multiple ways to measure the performance of our proposed models.
While we can directly compare the output angle values as a metric, 
we are more interested in the models learning to predict angles that result in close-to minimal energies.
For this we do zero-shot, fixed variable evaluation of the circuit.
In this evaluation we focus on comparing the result of the baseline energy 
with the energy that the respective predictor can achieve when directly applying its variables to the circuit.
Our goal is to determine and compare the model capability of producing angles, and thus states, 
close to, or better than, the ground state found by the VQE optimization.
The steps for evaluating a single molecule instance are the following:

\begin{enumerate}
    \item Use the perfect matching graph $G$ and the coordinates $C$ of the evaluated molecule in the forward pass of the model.
    \item Map the model predicted angles to the circuit ansatz.
    \item Simulate the resulting circuit and return the predicted energy, which is used for determining the model's accuracy.
\end{enumerate}

\subsection{Random Instance Evaluation}
For our first evaluation, we utilized the previously defined Algorithm \ref{alg:random-coordinates} for random data generation.
We generated 500 samples for each size $s \in \{2, 4, 6, 8, 10, 12\}$ of hydrogenic systems which we then run through the SPA optimization using the edge heuristic. The procedure is equivalent to the example code presented in the Appendix, including the orbital optimization step.

\begin{table}[ht]
  \caption{Comparison of fixed variables application methods against baseline, with the models generalizing up to H$_{12}$ from only learning on smaller instances. The exact amount of parameters are shown in \autoref{tab:models-overview}. The models were trained on two dataset sizes: large, i.e., 230k linear instances, and small, i.e., 1.000 linear instances of H$_4$ and 2.000 random instances of H$_6$, and evaluated on 500 instances of H$_{n}$.}
  \label{tab:fixed-method-comparison}
  \centering
  \resizebox{0.45\textwidth}{!}{
      \begin{tabular}{clccrr}
        \toprule
        Atoms & Method & Training size & Model size & ME (mEh) & MSE (mEh)\\
        \midrule
        \multirow{3}{*}{2}  & GAT & Large & Medium & 14 & 0.3 \\
         & Linear SchNet & Large & Small & 1 & 0.01\\
         & Mixed SchNet & Small & Large & 0.2 & $6 \times 10^{-4}$\\
        \midrule
        \multirow{3}{*}{4} & GAT & Large & Medium & 43 & 5 \\
         & Linear SchNet & Large & Small & 15 & 1 \\
         & Mixed SchNet & Small & Large & 6 & 0.3\\
        \midrule
        \multirow{3}{*}{6} & GAT & Large & Medium & 67 & 15  \\
         & Linear SchNet & Large & Small & 37 & 11 \\
         & Mixed SchNet & Small & Large & 10 & 2\\
        \midrule
        \multirow{3}{*}{8} & GAT & Large & Medium & 107 & 48 \\
         & Linear SchNet & Large & Small & 100 & 54 \\
         & Mixed SchNet & Small & Large & 34 & 5\\
        \midrule
        \multirow{3}{*}{10} & GAT & Large & Medium & 153 & 55  \\
         & Linear SchNet & Large & Small & 58 & 16 \\
         & Mixed SchNet & Small & Large & 26 & 4\\
        \midrule
        \multirow{3}{*}{12} & GAT & Large & Medium & 188 & 113 \\
         & Linear SchNet & Large & Small & 185 &  127 \\
         & Mixed SchNet & Small & Large & 38 & 6\\
        \bottomrule
      \end{tabular}
    }
\end{table}

The considered metrics are defined as follows:
\begin{equation}
  \begin{split}
    \text{MSE} &= \frac{\sum_{x \in S} (B_x - M_x)^2}{|S |} \\
    \text{ME} &= \frac{\sum_{x \in S} (B_x - M_x)}{|S|} \\
  \end{split}
\end{equation}
where $B_x$ is the baseline energy of the $x$-th sample, and $M_x$ is the corresponding method energy. 
The mean absolute error (MAE) is also often referred to as the $L_1$ loss metric, while the mean squared error (MSE) is commonly known as the $L_2$ loss metric. We are interested in the former since we do not want to lose information on the instances where $M_x < B_x$.
The final results for the fixed variable comparison can be seen in \autoref{tab:fixed-method-comparison}. The best model of each architecture is compared against the baseline, grouped by system size. We observe that for larger system sizes, the SchNet-based models perform better than the GATConv-based model, indicating comparatively best generalization capabilities provided by the architecture.
In particular, for the mixed SchNet model, the results indicate that small datasets are sufficient to capture the required information and predict ground state energies with adequate accuracy. The mixture of the dataset also has a positive effect for the generalizability where significant improvements where achieved when testing on randomized structures while the price of computing training instances of H$_6$ instead of H$_4$ was relatively small.

\begin{figure*}[ht]
  \begin{center}
    \includegraphics[width=0.95\textwidth]{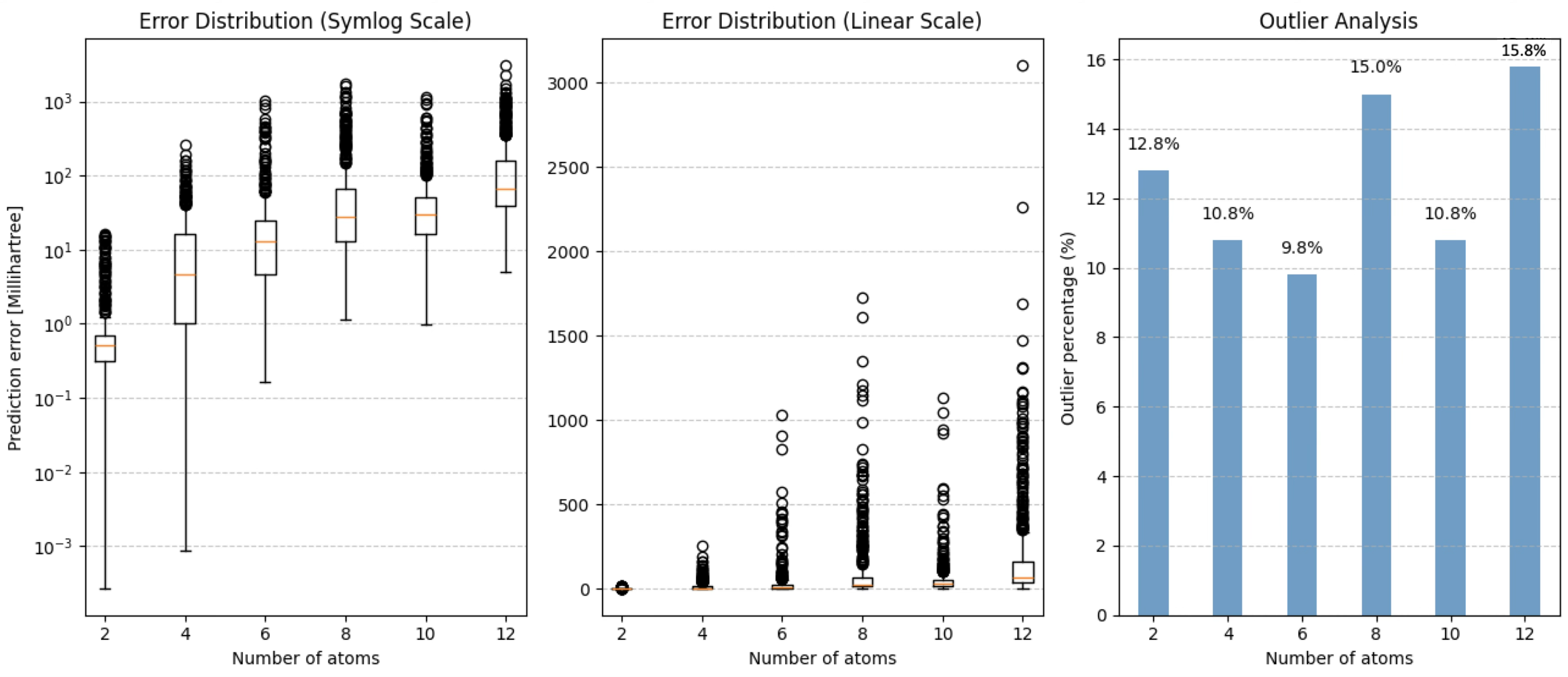}
  \end{center}
  \caption{Outliers analysis table for hydrogenic systems up to H$_{12}$. (Left-Center) Linear SchNet prediction error with respect to the baseline energy in symmetric-log and linear scale. (Right) Outlier frequency distribution for each molecule.}\label{fig:energydiff-baseline-schnet}
\end{figure*}

To gain more insight about the performance of our models, we carried out detailed comparisons of the energy differences.
\autoref{fig:energydiff-baseline-schnet} shows an outlier analysis, i.e., the respective count of samples that lie outside of the standard deviation, comparing the linear SchNet model and the baseline energy.
As expected, in our evaluation of the SchNet-based model, the largest amount of energy outliers occur for the H$_{12}$ system, i.e., the instances which differ the most from the linear H$_4$ training dataset and require the greatest transferability from the models. This is an indicator at potential improvements one can make to the model or dataset.

% now we look at linear
\subsection{Structured Instance Evaluation}
\begin{figure*}[ht]
  \begin{center}
    \subfigure[]{\includegraphics[width=0.49\textwidth]{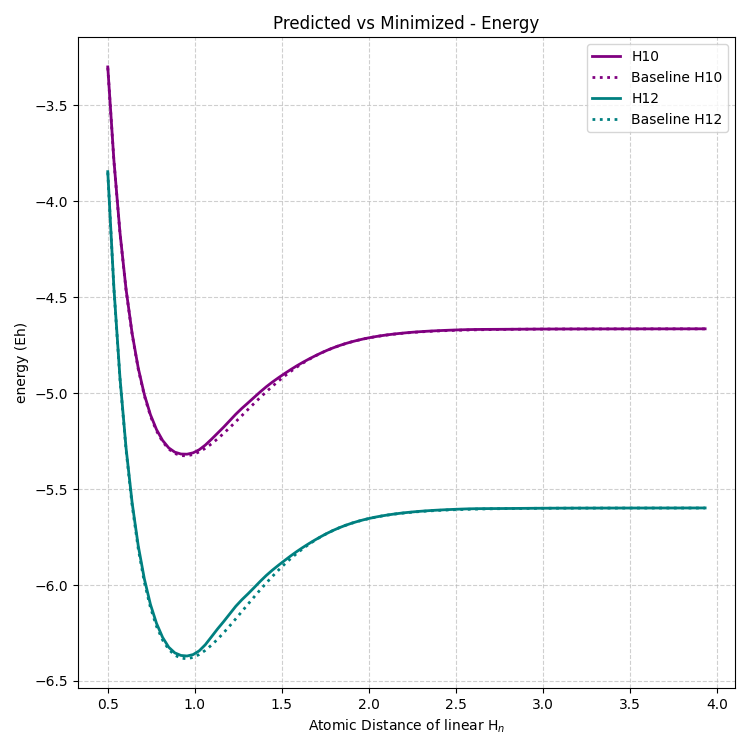}}
    \subfigure[]{\includegraphics[width=0.49\textwidth]{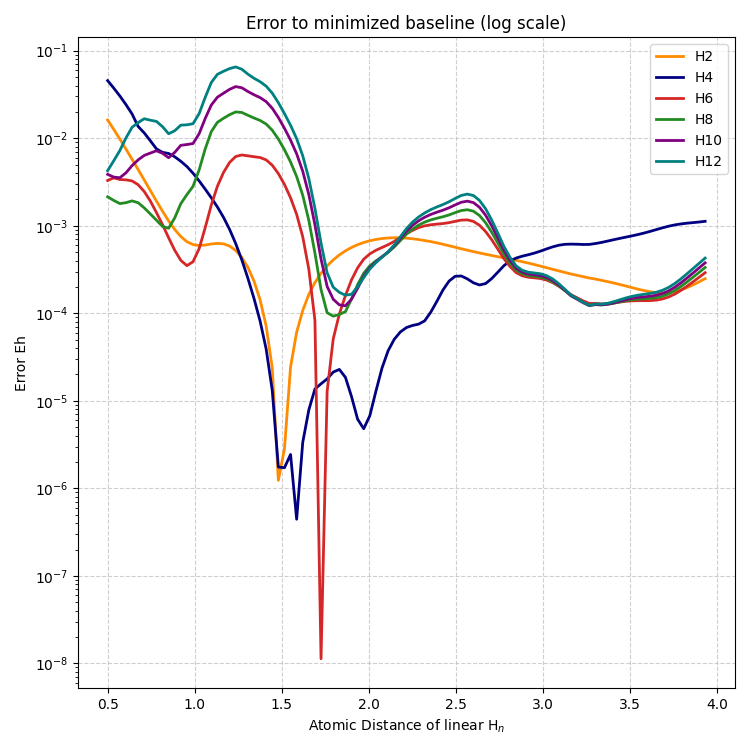}} \\
    \subfigure[]{\includegraphics[width=0.49\textwidth]{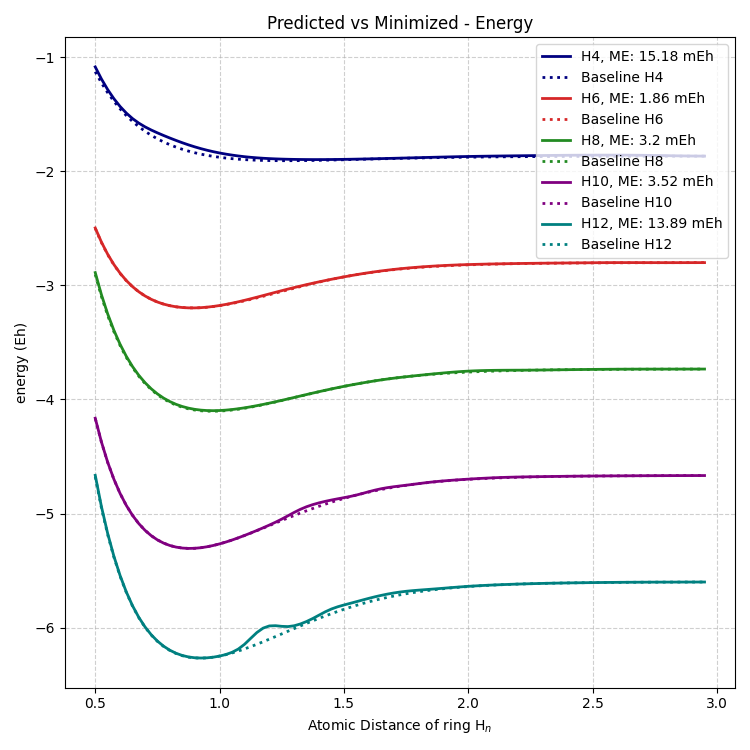}}
    \subfigure[]{\includegraphics[width=0.49\textwidth]{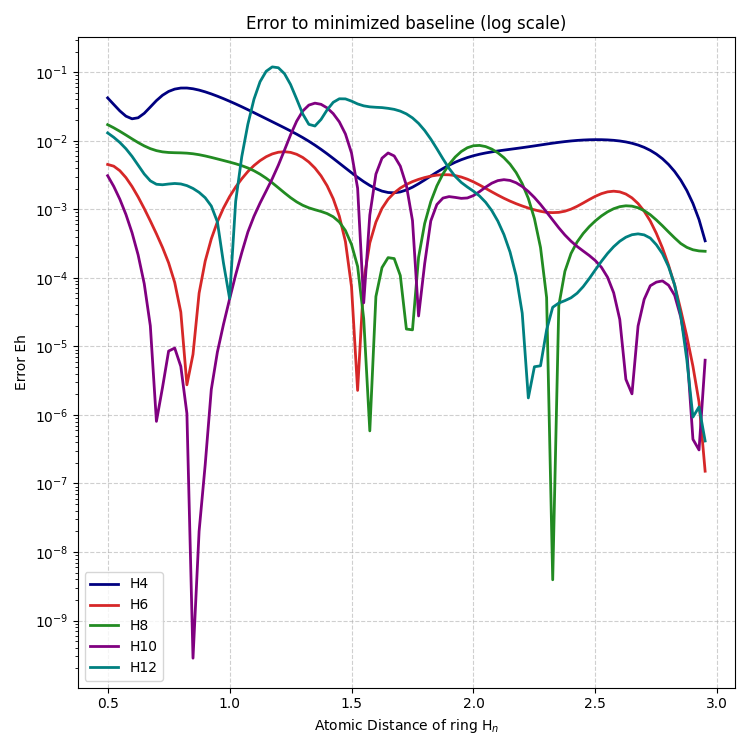}}
  \end{center}
  \caption{Evaluating linear SchNet model on linear and ring instances up to H$_{12}$, with the model being trained on 230k linear H$_4$ instances. Values are compared against the baseline SPA energy. (a) Model predictions for linear H$_{10}$ and H$_{12}$. (b) Model prediction errors on log-scale. (c) Model predictions for ring H$_n$. (d) Model prediction errors on log-scale.}\label{fig:linear_ring}
\end{figure*}

A further evaluation involves assessing model performance on structured instances. 
For this, we looked at the predicted ground state energies for linear and ring H$_n$ molecules, with the atom bonding lying on the x-axis ranging form 0.5 \AA \ to 4.0 \AA. The coordinates were generated using \autoref{alg:linear-coordinates} and \ref{alg:ring-coordinates}.
\autoref{fig:linear_ring} shows the errors for the linear SchNet model with a focus on the potential energy curves of the molecules.
The linear dataset is redundant since it is also partially in the training set, but it is a good proof of consistency that the SchNet-based models can transfer the learned structure to larger structured instances.
Using a log-scale for the plots makes it easier to see that, as expected, the model's accuracy drops as molecule size increases. Note however, that the prediction still remains physically sound while the chains and rings dissociate.

\begin{figure*}[ht]
  \begin{center}
    \subfigure[]{\includegraphics[width=0.49\textwidth]{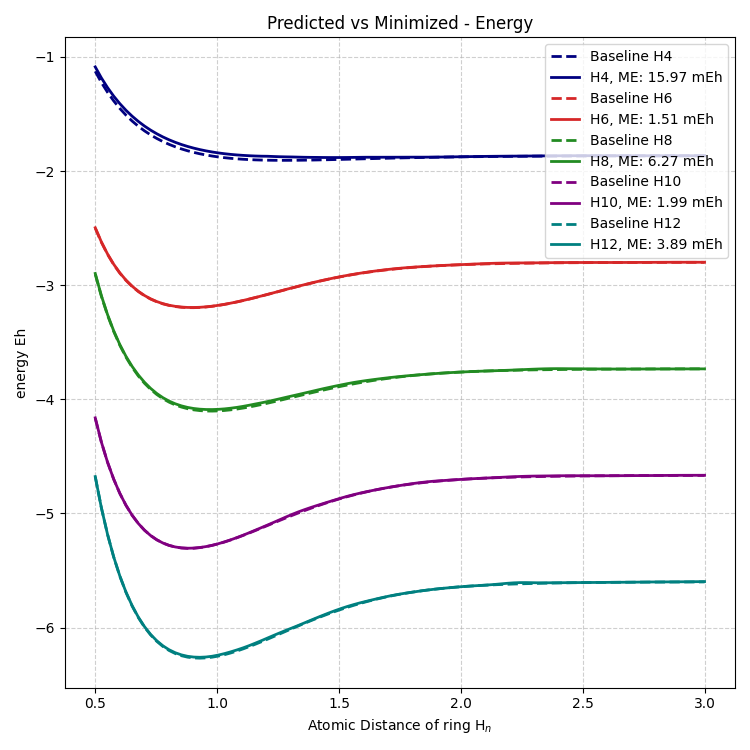}}
    \subfigure[]{\includegraphics[width=0.49\textwidth]{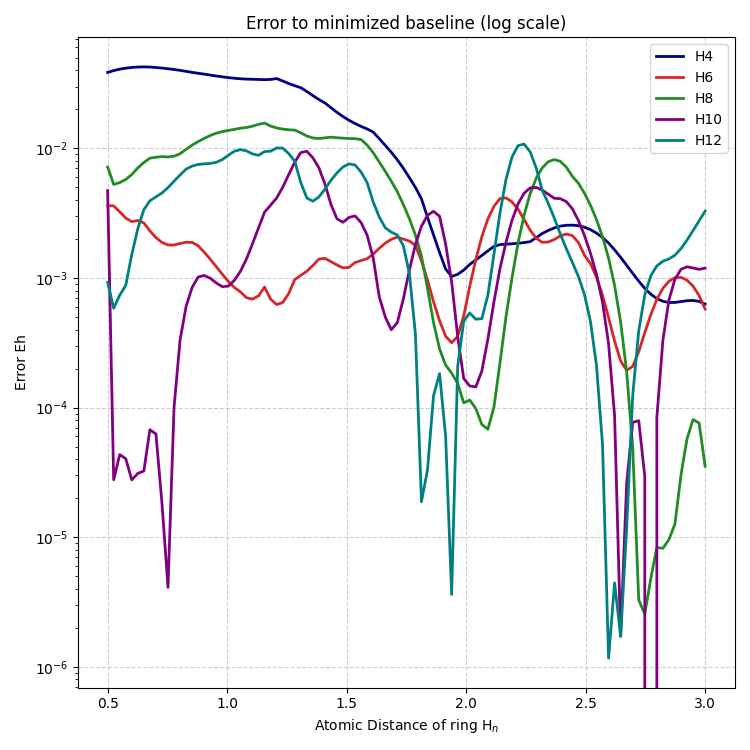}}
  \end{center}
  \caption{Evaluating mixed SchNet model on ring instances, generalizing up to H$_{12}$ instances, with the training dataset containing 1.000 linear instances of H$_4$ and 2.000 random instances of H$_6$. (a) Model predictions for ring H$_n$, compared against baseline SPA energy. (b) Model prediction errors on log-scale, compared against baseline SPA energy for H$_n$.}\label{fig:ring}
\end{figure*}

The mixed SchNet was evaluated on the ring instances in the same way with results shown in \autoref{fig:ring}.
Notably, the mixed model was able to generalize from linear and random instances to structured ring instances. Comparing this results with the linear SchNet ones shows that the accuracy improved by one order of magnitude, which is remarkable given the size of the training set.

\section{Conclusion and Outlook}
\label{c:conclusion}
In this work, we have explored the concept of modeling variational quantum parameters on SPA circuits, experimenting with three advanced approaches.
These have shown success in modeling the circuit parameters of hydrogenic systems, generalizing for unseen system sizes up to H$_{12}$. Notably, the forward pass of all three models requires only the coordinates and the edges of the best perfect matching graph for their predictions, showing that such fundamental data representation is sufficient to model molecular systems. In particular, the use of edges, which require a preprocessing step, show that perfect matching graphs accurately describe electronic correlation in a local manner.

The GAT and Linear SchNet models were trained on linear instances only, showing a good starting point for accurate results and generalizability to random and structured geometries. However, the proposed SchNet-based architectures proved best in predicting angles, yielding good variational parameters for the SPA ansatz. By further experimenting with such architecture, we have shown that by mixing the structure of the geometries in the training dataset we achieve a significant increase in accuracy. For this, we used linear and random molecule structures, showing that much of the necessary information can be extracted from such arrangements. This is true also for transferability, in fact, the model accurately predicts unseen instances, i.e., larger structured and random instances up to H$_{12}$.
Moreover, this last mixed model was trained with relatively small amount of data, demonstrating a better capability in capturing correlation information among atoms in the molecular structure.

Following this work, we propose multiple paths for further development.
Given larger and extended datasets and increased computational resources, 
we assume it is possible to enhance the models, improving on accuracy and transferability across various geometries. 
This could facilitate initialization of larger system sizes, as well as generalization to non-hydrogenic systems.
In addition, computational complexity can be increased not only through larger training datasets but also through greater quality in the VQE-based solutions.
This is due to the challenge in finding the global minimum of the optimization landscape. Finally, there are many possibilities for improving the models through alternative machine learning techniques, both in terms of architecture and learning strategies, such as semi-supervised learning or reinforcement learning.\\
A natural next step is to extend the most promising (mixed SchNet) approach of this work from minimal basis hydrogenic systems to (organic) molecules in more chemically accurate bases. Here we see the challenges on a rather technical side resulting from the many technicalities introduced by the quantum chemical methodology necessary for the task. Regarding the used chemical basis sets, one way to improve would be to switch to numerically optimized orbitals.~\cite{kottmann2020reducing, Langkabel2022} 
A grander challenge remains in adopting the methodology to more expressive circuit designs -- ideal candidates would be extended graph-based designs~\cite{kottmannMolecular2023}, localized hierarchical approaches~\cite{anselmetti2021local, burton2024accurate} or established coupled-cluster  method~\cite{lee2018generalized}. Note that all three contain the here used separable pair design as substructure.\\
We hope to provide a solid foundation for such future endeavors with this work.

\section{Acknowledgments}
This work has been funded by the Hightech Agenda Bayern and the Munich Quantum Valley through the Lighthouse project KID-QC$^2$. The authors thank Francisco Javier Del Arco Santos for various fruitful discussions.

\bibliography{main}

\begin{thebibliography}{52}%
\makeatletter
\providecommand \@ifxundefined [1]{%
 \@ifx{#1\undefined}
}%
\providecommand \@ifnum [1]{%
 \ifnum #1\expandafter \@firstoftwo
 \else \expandafter \@secondoftwo
 \fi
}%
\providecommand \@ifx [1]{%
 \ifx #1\expandafter \@firstoftwo
 \else \expandafter \@secondoftwo
 \fi
}%
\providecommand \natexlab [1]{#1}%
\providecommand \enquote  [1]{``#1''}%
\providecommand \bibnamefont  [1]{#1}%
\providecommand \bibfnamefont [1]{#1}%
\providecommand \citenamefont [1]{#1}%
\providecommand \href@noop [0]{\@secondoftwo}%
\providecommand \href [0]{\begingroup \@sanitize@url \@href}%
\providecommand \@href[1]{\@@startlink{#1}\@@href}%
\providecommand \@@href[1]{\endgroup#1\@@endlink}%
\providecommand \@sanitize@url [0]{\catcode `\\12\catcode `\$12\catcode
  `\&12\catcode `\#12\catcode `\^12\catcode `\_12\catcode `\%12\relax}%
\providecommand \@@startlink[1]{}%
\providecommand \@@endlink[0]{}%
\providecommand \url  [0]{\begingroup\@sanitize@url \@url }%
\providecommand \@url [1]{\endgroup\@href {#1}{\urlprefix }}%
\providecommand \urlprefix  [0]{URL }%
\providecommand \Eprint [0]{\href }%
\providecommand \doibase [0]{https://doi.org/}%
\providecommand \selectlanguage [0]{\@gobble}%
\providecommand \bibinfo  [0]{\@secondoftwo}%
\providecommand \bibfield  [0]{\@secondoftwo}%
\providecommand \translation [1]{[#1]}%
\providecommand \BibitemOpen [0]{}%
\providecommand \bibitemStop [0]{}%
\providecommand \bibitemNoStop [0]{.\EOS\space}%
\providecommand \EOS [0]{\spacefactor3000\relax}%
\providecommand \BibitemShut  [1]{\csname bibitem#1\endcsname}%
\let\auto@bib@innerbib\@empty
%</preamble>
\bibitem [{\citenamefont {McArdle}\ \emph {et~al.}(2020)\citenamefont
  {McArdle}, \citenamefont {Endo}, \citenamefont {Aspuru-Guzik}, \citenamefont
  {Benjamin},\ and\ \citenamefont {Yuan}}]{quantum-comp-chem}%
  \BibitemOpen
  \bibfield  {author} {\bibinfo {author} {\bibfnamefont {S.}~\bibnamefont
  {McArdle}}, \bibinfo {author} {\bibfnamefont {S.}~\bibnamefont {Endo}},
  \bibinfo {author} {\bibfnamefont {A.}~\bibnamefont {Aspuru-Guzik}}, \bibinfo
  {author} {\bibfnamefont {S.~C.}\ \bibnamefont {Benjamin}},\ and\ \bibinfo
  {author} {\bibfnamefont {X.}~\bibnamefont {Yuan}},\ }\bibfield  {title}
  {\bibinfo {title} {Quantum computational chemistry},\ }\bibfield  {journal}
  {\bibinfo  {journal} {Reviews of Modern Physics}\ }\textbf {\bibinfo {volume}
  {92}},\ \href {https://doi.org/10.1103/revmodphys.92.015003}
  {10.1103/revmodphys.92.015003} (\bibinfo {year} {2020})\BibitemShut {NoStop}%
\bibitem [{\citenamefont {{Aspuru-Guzik}}\ \emph {et~al.}(2005)\citenamefont
  {{Aspuru-Guzik}}, \citenamefont {Dutoi}, \citenamefont {Love},\ and\
  \citenamefont {{Head-Gordon}}}]{aspuru2005simulated}%
  \BibitemOpen
  \bibfield  {author} {\bibinfo {author} {\bibfnamefont {A.}~\bibnamefont
  {{Aspuru-Guzik}}}, \bibinfo {author} {\bibfnamefont {A.~D.}\ \bibnamefont
  {Dutoi}}, \bibinfo {author} {\bibfnamefont {P.~J.}\ \bibnamefont {Love}},\
  and\ \bibinfo {author} {\bibfnamefont {M.}~\bibnamefont {{Head-Gordon}}},\
  }\bibfield  {title} {\bibinfo {title} {Simulated quantum computation of
  molecular energies},\ }\href {https://doi.org/10.1126/science.1113479}
  {\bibfield  {journal} {\bibinfo  {journal} {Science (New York, N.Y.)}\
  }\textbf {\bibinfo {volume} {309}},\ \bibinfo {pages} {1704} (\bibinfo {year}
  {2005})}\BibitemShut {NoStop}%
\bibitem [{\citenamefont {Reiher}\ \emph {et~al.}(2017)\citenamefont {Reiher},
  \citenamefont {Wiebe}, \citenamefont {Svore}, \citenamefont {Wecker},\ and\
  \citenamefont {Troyer}}]{reiher2017elucidating}%
  \BibitemOpen
  \bibfield  {author} {\bibinfo {author} {\bibfnamefont {M.}~\bibnamefont
  {Reiher}}, \bibinfo {author} {\bibfnamefont {N.}~\bibnamefont {Wiebe}},
  \bibinfo {author} {\bibfnamefont {K.~M.}\ \bibnamefont {Svore}}, \bibinfo
  {author} {\bibfnamefont {D.}~\bibnamefont {Wecker}},\ and\ \bibinfo {author}
  {\bibfnamefont {M.}~\bibnamefont {Troyer}},\ }\bibfield  {title} {\bibinfo
  {title} {Elucidating reaction mechanisms on quantum computers},\ }\href
  {https://doi.org/10.1073/pnas.1619152114} {\bibfield  {journal} {\bibinfo
  {journal} {Proceedings of the National Academy of Sciences}\ }\textbf
  {\bibinfo {volume} {114}},\ \bibinfo {pages} {7555} (\bibinfo {year}
  {2017})},\ \Eprint
  {https://arxiv.org/abs/https://www.pnas.org/content/114/29/7555.full.pdf}
  {https://www.pnas.org/content/114/29/7555.full.pdf} \BibitemShut {NoStop}%
\bibitem [{\citenamefont {{von Burg}}\ \emph {et~al.}(2021)\citenamefont {{von
  Burg}}, \citenamefont {Low}, \citenamefont {H{\"a}ner}, \citenamefont
  {Steiger}, \citenamefont {Reiher}, \citenamefont {Roetteler},\ and\
  \citenamefont {Troyer}}]{vonburg2021quantum}%
  \BibitemOpen
  \bibfield  {author} {\bibinfo {author} {\bibfnamefont {V.}~\bibnamefont {{von
  Burg}}}, \bibinfo {author} {\bibfnamefont {G.~H.}\ \bibnamefont {Low}},
  \bibinfo {author} {\bibfnamefont {T.}~\bibnamefont {H{\"a}ner}}, \bibinfo
  {author} {\bibfnamefont {D.~S.}\ \bibnamefont {Steiger}}, \bibinfo {author}
  {\bibfnamefont {M.}~\bibnamefont {Reiher}}, \bibinfo {author} {\bibfnamefont
  {M.}~\bibnamefont {Roetteler}},\ and\ \bibinfo {author} {\bibfnamefont
  {M.}~\bibnamefont {Troyer}},\ }\bibfield  {title} {\bibinfo {title} {Quantum
  computing enhanced computational catalysis},\ }\href@noop {} {\  (\bibinfo
  {year} {2021})},\ \Eprint {https://arxiv.org/abs/2007.14460}
  {arxiv:2007.14460 [quant-ph]} \BibitemShut {NoStop}%
\bibitem [{\citenamefont {Rocca}\ \emph {et~al.}(2024)\citenamefont {Rocca},
  \citenamefont {Cortes}, \citenamefont {Gonthier}, \citenamefont {Ollitrault},
  \citenamefont {Parrish}, \citenamefont {Anselmetti}, \citenamefont {others},\
  and\ \citenamefont {Streif}}]{rocca2024reducing}%
  \BibitemOpen
  \bibfield  {author} {\bibinfo {author} {\bibfnamefont {D.}~\bibnamefont
  {Rocca}}, \bibinfo {author} {\bibfnamefont {C.~L.}\ \bibnamefont {Cortes}},
  \bibinfo {author} {\bibfnamefont {J.~F.}\ \bibnamefont {Gonthier}}, \bibinfo
  {author} {\bibfnamefont {P.~J.}\ \bibnamefont {Ollitrault}}, \bibinfo
  {author} {\bibfnamefont {R.~M.}\ \bibnamefont {Parrish}}, \bibinfo {author}
  {\bibfnamefont {G.~L.}\ \bibnamefont {Anselmetti}}, \bibinfo {author}
  {\bibnamefont {others}},\ and\ \bibinfo {author} {\bibfnamefont
  {M.}~\bibnamefont {Streif}},\ }\bibfield  {title} {\bibinfo {title} {Reducing
  the runtime of fault-tolerant quantum simulations in chemistry through
  symmetry-compressed double factorization},\ }\href@noop {} {\bibfield
  {journal} {\bibinfo  {journal} {Journal of Chemical Theory and Computation}\
  }\textbf {\bibinfo {volume} {20}},\ \bibinfo {pages} {4639} (\bibinfo {year}
  {2024})}\BibitemShut {NoStop}%
\bibitem [{\citenamefont {Cortes}\ \emph {et~al.}(2024)\citenamefont {Cortes},
  \citenamefont {Rocca}, \citenamefont {Gonthier}, \citenamefont {Ollitrault},
  \citenamefont {Parrish}, \citenamefont {Anselmetti}, \citenamefont {others},\
  and\ \citenamefont {Streif}}]{cortes2024assessing}%
  \BibitemOpen
  \bibfield  {author} {\bibinfo {author} {\bibfnamefont {C.~L.}\ \bibnamefont
  {Cortes}}, \bibinfo {author} {\bibfnamefont {D.}~\bibnamefont {Rocca}},
  \bibinfo {author} {\bibfnamefont {J.~F.}\ \bibnamefont {Gonthier}}, \bibinfo
  {author} {\bibfnamefont {P.~J.}\ \bibnamefont {Ollitrault}}, \bibinfo
  {author} {\bibfnamefont {R.~M.}\ \bibnamefont {Parrish}}, \bibinfo {author}
  {\bibfnamefont {G.~L.~R.}\ \bibnamefont {Anselmetti}}, \bibinfo {author}
  {\bibnamefont {others}},\ and\ \bibinfo {author} {\bibfnamefont
  {M.}~\bibnamefont {Streif}},\ }\bibfield  {title} {\bibinfo {title}
  {Assessing the query complexity limits of quantum phase estimation using
  symmetry-aware spectral bounds},\ }\href@noop {} {\bibfield  {journal}
  {\bibinfo  {journal} {Physical Review A}\ }\textbf {\bibinfo {volume}
  {110}},\ \bibinfo {pages} {022420} (\bibinfo {year} {2024})}\BibitemShut
  {NoStop}%
\bibitem [{\citenamefont {Ollitrault}\ \emph {et~al.}(2025)\citenamefont
  {Ollitrault}, \citenamefont {Gonthier}, \citenamefont {Rocca}, \citenamefont
  {Anselmetti}, \citenamefont {Degroote}, \citenamefont {Moll}, \citenamefont
  {others},\ and\ \citenamefont {Streif}}]{ollitrault2025improving}%
  \BibitemOpen
  \bibfield  {author} {\bibinfo {author} {\bibfnamefont {P.~J.}\ \bibnamefont
  {Ollitrault}}, \bibinfo {author} {\bibfnamefont {J.~F.}\ \bibnamefont
  {Gonthier}}, \bibinfo {author} {\bibfnamefont {D.}~\bibnamefont {Rocca}},
  \bibinfo {author} {\bibfnamefont {G.~L.}\ \bibnamefont {Anselmetti}},
  \bibinfo {author} {\bibfnamefont {M.}~\bibnamefont {Degroote}}, \bibinfo
  {author} {\bibfnamefont {N.}~\bibnamefont {Moll}}, \bibinfo {author}
  {\bibnamefont {others}},\ and\ \bibinfo {author} {\bibfnamefont
  {M.}~\bibnamefont {Streif}},\ }\bibfield  {title} {\bibinfo {title}
  {Improving the runtime of quantum phase estimation for chemistry through
  basis set optimization},\ }\href@noop {} {\bibfield  {journal} {\bibinfo
  {journal} {arXiv preprint arXiv:2509.05733}\ } (\bibinfo {year}
  {2025})}\BibitemShut {NoStop}%
\bibitem [{\citenamefont {Stroschein}\ \emph {et~al.}(2025)\citenamefont
  {Stroschein}, \citenamefont {Castaldo},\ and\ \citenamefont
  {Reiher}}]{stroschein2025groundexcitedstateenergiesanalytic}%
  \BibitemOpen
  \bibfield  {author} {\bibinfo {author} {\bibfnamefont {T.}~\bibnamefont
  {Stroschein}}, \bibinfo {author} {\bibfnamefont {D.}~\bibnamefont
  {Castaldo}},\ and\ \bibinfo {author} {\bibfnamefont {M.}~\bibnamefont
  {Reiher}},\ }\bibfield  {title} {\bibinfo {title} {Ground and excited-state
  energies with analytic errors and short time evolution on a quantum
  computer},\ }\href {https://arxiv.org/abs/2507.15148} {\  (\bibinfo {year}
  {2025})},\ \Eprint {https://arxiv.org/abs/2507.15148} {arXiv:2507.15148
  [quant-ph]} \BibitemShut {NoStop}%
\bibitem [{\citenamefont {Motta}\ \emph {et~al.}(2020)\citenamefont {Motta},
  \citenamefont {Sun}, \citenamefont {Tan}, \citenamefont {O'Rourke},
  \citenamefont {Ye}, \citenamefont {Minnich}, \citenamefont {Brand{\~a}o},\
  and\ \citenamefont {Chan}}]{motta2020determining}%
  \BibitemOpen
  \bibfield  {author} {\bibinfo {author} {\bibfnamefont {M.}~\bibnamefont
  {Motta}}, \bibinfo {author} {\bibfnamefont {C.}~\bibnamefont {Sun}}, \bibinfo
  {author} {\bibfnamefont {A.~T.}\ \bibnamefont {Tan}}, \bibinfo {author}
  {\bibfnamefont {M.~J.}\ \bibnamefont {O'Rourke}}, \bibinfo {author}
  {\bibfnamefont {E.}~\bibnamefont {Ye}}, \bibinfo {author} {\bibfnamefont
  {A.~J.}\ \bibnamefont {Minnich}}, \bibinfo {author} {\bibfnamefont {F.~G.}\
  \bibnamefont {Brand{\~a}o}},\ and\ \bibinfo {author} {\bibfnamefont
  {G.~K.-L.}\ \bibnamefont {Chan}},\ }\bibfield  {title} {\bibinfo {title}
  {Determining eigenstates and thermal states on a quantum computer using
  quantum imaginary time evolution},\ }\href
  {https://doi.org/10.1038/s41567-019-0704-4} {\bibfield  {journal} {\bibinfo
  {journal} {Nature Physics}\ }\textbf {\bibinfo {volume} {16}},\ \bibinfo
  {pages} {205} (\bibinfo {year} {2020})}\BibitemShut {NoStop}%
\bibitem [{\citenamefont {Langkabel}\ and\ \citenamefont
  {Bande}(2022)}]{Langkabel2022}%
  \BibitemOpen
  \bibfield  {author} {\bibinfo {author} {\bibfnamefont {F.}~\bibnamefont
  {Langkabel}}\ and\ \bibinfo {author} {\bibfnamefont {A.}~\bibnamefont
  {Bande}},\ }\bibfield  {title} {\bibinfo {title} {Quantum-compute algorithm
  for exact laser-driven electron dynamics in molecules},\ }\href
  {https://doi.org/10.1021/acs.jctc.2c00701} {\bibfield  {journal} {\bibinfo
  {journal} {Journal of Chemical Theory and Computation}\ }\textbf {\bibinfo
  {volume} {18}},\ \bibinfo {pages} {7082} (\bibinfo {year}
  {2022})}\BibitemShut {NoStop}%
\bibitem [{\citenamefont {Peruzzo}\ \emph {et~al.}(2014)\citenamefont
  {Peruzzo}, \citenamefont {McClean}, \citenamefont {Shadbolt}, \citenamefont
  {Yung}, \citenamefont {Zhou}, \citenamefont {Love}, \citenamefont
  {{Aspuru-Guzik}},\ and\ \citenamefont {O'brien}}]{peruzzo2014variational}%
  \BibitemOpen
  \bibfield  {author} {\bibinfo {author} {\bibfnamefont {A.}~\bibnamefont
  {Peruzzo}}, \bibinfo {author} {\bibfnamefont {J.}~\bibnamefont {McClean}},
  \bibinfo {author} {\bibfnamefont {P.}~\bibnamefont {Shadbolt}}, \bibinfo
  {author} {\bibfnamefont {M.-H.}\ \bibnamefont {Yung}}, \bibinfo {author}
  {\bibfnamefont {X.-Q.}\ \bibnamefont {Zhou}}, \bibinfo {author}
  {\bibfnamefont {P.~J.}\ \bibnamefont {Love}}, \bibinfo {author}
  {\bibfnamefont {A.}~\bibnamefont {{Aspuru-Guzik}}},\ and\ \bibinfo {author}
  {\bibfnamefont {J.~L.}\ \bibnamefont {O'brien}},\ }\bibfield  {title}
  {\bibinfo {title} {A variational eigenvalue solver on a photonic quantum
  processor},\ }\href {https://doi.org/10.1038/ncomms5213} {\bibfield
  {journal} {\bibinfo  {journal} {Nature Communications}\ }\textbf {\bibinfo
  {volume} {5}},\ \bibinfo {pages} {4213} (\bibinfo {year} {2014})}\BibitemShut
  {NoStop}%
\bibitem [{\citenamefont {Anand}\ \emph {et~al.}(2022)\citenamefont {Anand},
  \citenamefont {Schleich}, \citenamefont {{Alperin-Lea}}, \citenamefont
  {Jensen}, \citenamefont {Sim}, \citenamefont {{D{\'i}az-Tinoco}},
  \citenamefont {Kottmann}, \citenamefont {Degroote}, \citenamefont
  {Izmaylov},\ and\ \citenamefont {{Aspuru-Guzik}}}]{anand2022quantum}%
  \BibitemOpen
  \bibfield  {author} {\bibinfo {author} {\bibfnamefont {A.}~\bibnamefont
  {Anand}}, \bibinfo {author} {\bibfnamefont {P.}~\bibnamefont {Schleich}},
  \bibinfo {author} {\bibfnamefont {S.}~\bibnamefont {{Alperin-Lea}}}, \bibinfo
  {author} {\bibfnamefont {P.~W.~K.}\ \bibnamefont {Jensen}}, \bibinfo {author}
  {\bibfnamefont {S.}~\bibnamefont {Sim}}, \bibinfo {author} {\bibfnamefont
  {M.}~\bibnamefont {{D{\'i}az-Tinoco}}}, \bibinfo {author} {\bibfnamefont
  {J.~S.}\ \bibnamefont {Kottmann}}, \bibinfo {author} {\bibfnamefont
  {M.}~\bibnamefont {Degroote}}, \bibinfo {author} {\bibfnamefont {A.~F.}\
  \bibnamefont {Izmaylov}},\ and\ \bibinfo {author} {\bibfnamefont
  {A.}~\bibnamefont {{Aspuru-Guzik}}},\ }\bibfield  {title} {\bibinfo {title}
  {A quantum computing view on unitary coupled cluster theory},\ }\href
  {https://doi.org/10.1039/D1CS00932J} {\bibfield  {journal} {\bibinfo
  {journal} {Chemical Society Reviews}\ }\textbf {\bibinfo {volume} {51}},\
  \bibinfo {pages} {1659} (\bibinfo {year} {2022})}\BibitemShut {NoStop}%
\bibitem [{\citenamefont {Tilly}\ \emph {et~al.}(2022)\citenamefont {Tilly},
  \citenamefont {Chen}, \citenamefont {Cao}, \citenamefont {Picozzi},
  \citenamefont {Setia}, \citenamefont {Li}, \citenamefont {Grant},
  \citenamefont {Wossnig}, \citenamefont {Rungger}, \citenamefont {Booth},\
  and\ \citenamefont {Tennyson}}]{tillyVariationalQuantumEigensolver2022}%
  \BibitemOpen
  \bibfield  {author} {\bibinfo {author} {\bibfnamefont {J.}~\bibnamefont
  {Tilly}}, \bibinfo {author} {\bibfnamefont {H.}~\bibnamefont {Chen}},
  \bibinfo {author} {\bibfnamefont {S.}~\bibnamefont {Cao}}, \bibinfo {author}
  {\bibfnamefont {D.}~\bibnamefont {Picozzi}}, \bibinfo {author} {\bibfnamefont
  {K.}~\bibnamefont {Setia}}, \bibinfo {author} {\bibfnamefont
  {Y.}~\bibnamefont {Li}}, \bibinfo {author} {\bibfnamefont {E.}~\bibnamefont
  {Grant}}, \bibinfo {author} {\bibfnamefont {L.}~\bibnamefont {Wossnig}},
  \bibinfo {author} {\bibfnamefont {I.}~\bibnamefont {Rungger}}, \bibinfo
  {author} {\bibfnamefont {G.~H.}\ \bibnamefont {Booth}},\ and\ \bibinfo
  {author} {\bibfnamefont {J.}~\bibnamefont {Tennyson}},\ }\bibfield  {title}
  {\bibinfo {title} {The {{Variational Quantum Eigensolver}}: {{A}} review of
  methods and best practices},\ }\href
  {https://doi.org/10.1016/j.physrep.2022.08.003} {\bibfield  {journal}
  {\bibinfo  {journal} {Physics Reports}\ }\bibinfo {series} {The {{Variational
  Quantum Eigensolver}}: A Review of Methods and Best Practices},\ \textbf
  {\bibinfo {volume} {986}},\ \bibinfo {pages} {1} (\bibinfo {year}
  {2022})}\BibitemShut {NoStop}%
\bibitem [{\citenamefont {Cerezo}\ \emph {et~al.}(2021)\citenamefont {Cerezo},
  \citenamefont {Arrasmith}, \citenamefont {Babbush}, \citenamefont {Benjamin},
  \citenamefont {Endo}, \citenamefont {Fujii}, \citenamefont {McClean},
  \citenamefont {Mitarai}, \citenamefont {Yuan}, \citenamefont {Cincio}, ,\
  and\ \citenamefont {Coles}}]{cerezo2021variational}%
  \BibitemOpen
  \bibfield  {author} {\bibinfo {author} {\bibfnamefont {M.}~\bibnamefont
  {Cerezo}}, \bibinfo {author} {\bibfnamefont {A.}~\bibnamefont {Arrasmith}},
  \bibinfo {author} {\bibfnamefont {R.}~\bibnamefont {Babbush}}, \bibinfo
  {author} {\bibfnamefont {S.~C.}\ \bibnamefont {Benjamin}}, \bibinfo {author}
  {\bibfnamefont {S.}~\bibnamefont {Endo}}, \bibinfo {author} {\bibfnamefont
  {K.}~\bibnamefont {Fujii}}, \bibinfo {author} {\bibfnamefont {J.~R.}\
  \bibnamefont {McClean}}, \bibinfo {author} {\bibfnamefont {K.}~\bibnamefont
  {Mitarai}}, \bibinfo {author} {\bibfnamefont {X.}~\bibnamefont {Yuan}},
  \bibinfo {author} {\bibfnamefont {L.}~\bibnamefont {Cincio}}, ,\ and\
  \bibinfo {author} {\bibfnamefont {P.~J.}\ \bibnamefont {Coles}},\ }\bibfield
  {title} {\bibinfo {title} {Variational quantum algorithms},\ }\href
  {https://doi.org/10.1038/s42254-021-00348-9} {\bibfield  {journal} {\bibinfo
  {journal} {Nature Reviews Physics}\ }\textbf {\bibinfo {volume} {3}},\
  \bibinfo {pages} {625} (\bibinfo {year} {2021})}\BibitemShut {NoStop}%
\bibitem [{\citenamefont {Bharti}\ \emph {et~al.}(2022)\citenamefont {Bharti},
  \citenamefont {{Cervera-Lierta}}, \citenamefont {Kyaw}, \citenamefont {Haug},
  \citenamefont {{Alperin-Lea}}, \citenamefont {Anand}, \citenamefont
  {Degroote}, \citenamefont {Heimonen}, \citenamefont {Kottmann}, \citenamefont
  {Menke}, \citenamefont {Mok}, \citenamefont {Sim}, \citenamefont {Kwek},\
  and\ \citenamefont {Aspuru-Guzik}}]{bharti2022noisy}%
  \BibitemOpen
  \bibfield  {author} {\bibinfo {author} {\bibfnamefont {K.}~\bibnamefont
  {Bharti}}, \bibinfo {author} {\bibfnamefont {A.}~\bibnamefont
  {{Cervera-Lierta}}}, \bibinfo {author} {\bibfnamefont {T.~H.}\ \bibnamefont
  {Kyaw}}, \bibinfo {author} {\bibfnamefont {T.}~\bibnamefont {Haug}}, \bibinfo
  {author} {\bibfnamefont {S.}~\bibnamefont {{Alperin-Lea}}}, \bibinfo {author}
  {\bibfnamefont {A.}~\bibnamefont {Anand}}, \bibinfo {author} {\bibfnamefont
  {M.}~\bibnamefont {Degroote}}, \bibinfo {author} {\bibfnamefont
  {H.}~\bibnamefont {Heimonen}}, \bibinfo {author} {\bibfnamefont {J.~S.}\
  \bibnamefont {Kottmann}}, \bibinfo {author} {\bibfnamefont {T.}~\bibnamefont
  {Menke}}, \bibinfo {author} {\bibfnamefont {W.-K.}\ \bibnamefont {Mok}},
  \bibinfo {author} {\bibfnamefont {S.}~\bibnamefont {Sim}}, \bibinfo {author}
  {\bibfnamefont {L.-C.}\ \bibnamefont {Kwek}},\ and\ \bibinfo {author}
  {\bibfnamefont {A.}~\bibnamefont {Aspuru-Guzik}},\ }\bibfield  {title}
  {\bibinfo {title} {Noisy intermediate-scale quantum algorithms},\ }\href
  {https://doi.org/10.1103/RevModPhys.94.015004} {\bibfield  {journal}
  {\bibinfo  {journal} {Reviews of Modern Physics}\ }\textbf {\bibinfo {volume}
  {94}},\ \bibinfo {pages} {015004} (\bibinfo {year} {2022})}\BibitemShut
  {NoStop}%
\bibitem [{\citenamefont {Romero}\ \emph {et~al.}(2018)\citenamefont {Romero},
  \citenamefont {Babbush}, \citenamefont {McClean}, \citenamefont {Hempel},
  \citenamefont {Love},\ and\ \citenamefont
  {{Aspuru-Guzik}}}]{romero2018strategies}%
  \BibitemOpen
  \bibfield  {author} {\bibinfo {author} {\bibfnamefont {J.}~\bibnamefont
  {Romero}}, \bibinfo {author} {\bibfnamefont {R.}~\bibnamefont {Babbush}},
  \bibinfo {author} {\bibfnamefont {J.~R.}\ \bibnamefont {McClean}}, \bibinfo
  {author} {\bibfnamefont {C.}~\bibnamefont {Hempel}}, \bibinfo {author}
  {\bibfnamefont {P.~J.}\ \bibnamefont {Love}},\ and\ \bibinfo {author}
  {\bibfnamefont {A.}~\bibnamefont {{Aspuru-Guzik}}},\ }\bibfield  {title}
  {\bibinfo {title} {Strategies for quantum computing molecular energies using
  the unitary coupled cluster ansatz},\ }\href
  {https://doi.org/10.1088/2058-9565/aad3e4} {\bibfield  {journal} {\bibinfo
  {journal} {Quantum Science and Technology}\ }\textbf {\bibinfo {volume}
  {4}},\ \bibinfo {pages} {014008} (\bibinfo {year} {2018})}\BibitemShut
  {NoStop}%
\bibitem [{\citenamefont {Lee}\ \emph {et~al.}(2018)\citenamefont {Lee},
  \citenamefont {Huggins}, \citenamefont {{Head-Gordon}},\ and\ \citenamefont
  {Whaley}}]{lee2018generalized}%
  \BibitemOpen
  \bibfield  {author} {\bibinfo {author} {\bibfnamefont {J.}~\bibnamefont
  {Lee}}, \bibinfo {author} {\bibfnamefont {W.~J.}\ \bibnamefont {Huggins}},
  \bibinfo {author} {\bibfnamefont {M.}~\bibnamefont {{Head-Gordon}}},\ and\
  \bibinfo {author} {\bibfnamefont {K.~B.}\ \bibnamefont {Whaley}},\ }\bibfield
   {title} {\bibinfo {title} {Generalized unitary coupled cluster wave
  functions for quantum computation},\ }\href
  {https://doi.org/10.1021/acs.jctc.8b01004} {\bibfield  {journal} {\bibinfo
  {journal} {Journal of Chemical Theory and Computation}\ }\textbf {\bibinfo
  {volume} {15}},\ \bibinfo {pages} {311} (\bibinfo {year} {2018})}\BibitemShut
  {NoStop}%
\bibitem [{\citenamefont {Kottmann}\ and\ \citenamefont
  {{Aspuru-Guzik}}(2022)}]{kottmann2022optimized}%
  \BibitemOpen
  \bibfield  {author} {\bibinfo {author} {\bibfnamefont {J.~S.}\ \bibnamefont
  {Kottmann}}\ and\ \bibinfo {author} {\bibfnamefont {A.}~\bibnamefont
  {{Aspuru-Guzik}}},\ }\bibfield  {title} {\bibinfo {title} {Optimized
  low-depth quantum circuits for molecular electronic structure using a
  separable-pair approximation},\ }\href
  {https://doi.org/10.1103/PhysRevA.105.032449} {\bibfield  {journal} {\bibinfo
   {journal} {Physical Review A}\ }\textbf {\bibinfo {volume} {105}},\ \bibinfo
  {pages} {032449} (\bibinfo {year} {2022})}\BibitemShut {NoStop}%
\bibitem [{\citenamefont {Kottmann}()}]{kottmannMolecular2023}%
  \BibitemOpen
  \bibfield  {author} {\bibinfo {author} {\bibfnamefont {J.~S.}\ \bibnamefont
  {Kottmann}},\ }\bibfield  {title} {\bibinfo {title} {Molecular {{Quantum
  Circuit Design}}: {{A Graph-Based Approach}}},\ }\href
  {https://doi.org/10.22331/q-2023-08-03-1073} {\bibfield  {journal} {\bibinfo
  {journal} {Quantum}\ }\textbf {\bibinfo {volume} {7}},\ \bibinfo {pages}
  {1073}}\BibitemShut {NoStop}%
\bibitem [{\citenamefont {Burton}\ \emph {et~al.}(2022)\citenamefont {Burton},
  \citenamefont {{Marti-Dafcik}}, \citenamefont {Tew},\ and\ \citenamefont
  {Wales}}]{burton2022exact}%
  \BibitemOpen
  \bibfield  {author} {\bibinfo {author} {\bibfnamefont {H.~G.~A.}\
  \bibnamefont {Burton}}, \bibinfo {author} {\bibfnamefont {D.}~\bibnamefont
  {{Marti-Dafcik}}}, \bibinfo {author} {\bibfnamefont {D.~P.}\ \bibnamefont
  {Tew}},\ and\ \bibinfo {author} {\bibfnamefont {D.~J.}\ \bibnamefont
  {Wales}},\ }\bibfield  {title} {\bibinfo {title} {Exact electronic states
  with shallow quantum circuits through global optimisation}\ }\href
  {https://doi.org/10.48550/arxiv.2207.00085} {10.48550/arxiv.2207.00085}
  (\bibinfo {year} {2022})\BibitemShut {NoStop}%
\bibitem [{\citenamefont {Burton}(2024)}]{burton2024accurate}%
  \BibitemOpen
  \bibfield  {author} {\bibinfo {author} {\bibfnamefont {H.~G.}\ \bibnamefont
  {Burton}},\ }\bibfield  {title} {\bibinfo {title} {Accurate and
  gate-efficient quantum ans{\"a}tze for electronic states without adaptive
  optimization},\ }\href@noop {} {\bibfield  {journal} {\bibinfo  {journal}
  {Physical Review Research}\ }\textbf {\bibinfo {volume} {6}},\ \bibinfo
  {pages} {023300} (\bibinfo {year} {2024})}\BibitemShut {NoStop}%
\bibitem [{\citenamefont {Weber}\ \emph {et~al.}(2022)\citenamefont {Weber},
  \citenamefont {Anand}, \citenamefont {Cervera-Lierta}, \citenamefont
  {Kottmann}, \citenamefont {Kyaw}, \citenamefont {Li}, \citenamefont
  {Aspuru-Guzik}, \citenamefont {Zhang},\ and\ \citenamefont
  {Zhao}}]{weber2022toward}%
  \BibitemOpen
  \bibfield  {author} {\bibinfo {author} {\bibfnamefont {M.}~\bibnamefont
  {Weber}}, \bibinfo {author} {\bibfnamefont {A.}~\bibnamefont {Anand}},
  \bibinfo {author} {\bibfnamefont {A.}~\bibnamefont {Cervera-Lierta}},
  \bibinfo {author} {\bibfnamefont {J.~S.}\ \bibnamefont {Kottmann}}, \bibinfo
  {author} {\bibfnamefont {T.~H.}\ \bibnamefont {Kyaw}}, \bibinfo {author}
  {\bibfnamefont {B.}~\bibnamefont {Li}}, \bibinfo {author} {\bibfnamefont
  {A.}~\bibnamefont {Aspuru-Guzik}}, \bibinfo {author} {\bibfnamefont
  {C.}~\bibnamefont {Zhang}},\ and\ \bibinfo {author} {\bibfnamefont
  {Z.}~\bibnamefont {Zhao}},\ }\bibfield  {title} {\bibinfo {title} {Toward
  reliability in the nisq era: Robust interval guarantee for quantum
  measurements on approximate states},\ }\href@noop {} {\bibfield  {journal}
  {\bibinfo  {journal} {Physical Review Research}\ }\textbf {\bibinfo {volume}
  {4}},\ \bibinfo {pages} {033217} (\bibinfo {year} {2022})}\BibitemShut
  {NoStop}%
\bibitem [{\citenamefont {Schleich}\ \emph {et~al.}(2021)\citenamefont
  {Schleich}, \citenamefont {Kottmann},\ and\ \citenamefont
  {{Aspuru-Guzik}}}]{schleich2021improving}%
  \BibitemOpen
  \bibfield  {author} {\bibinfo {author} {\bibfnamefont {P.}~\bibnamefont
  {Schleich}}, \bibinfo {author} {\bibfnamefont {J.~S.}\ \bibnamefont
  {Kottmann}},\ and\ \bibinfo {author} {\bibfnamefont {A.}~\bibnamefont
  {{Aspuru-Guzik}}},\ }\bibfield  {title} {\bibinfo {title} {Improving the
  accuracy of the variational quantum eigensolver for molecular systems by the
  explicitly-correlated perturbative [2]{{R12-Correction}}},\ }\href@noop {} {\
   (\bibinfo {year} {2021})},\ \Eprint {https://arxiv.org/abs/2110.06812}
  {arxiv:2110.06812 [quant-ph]} \BibitemShut {NoStop}%
\bibitem [{\citenamefont {Gil}\ \emph {et~al.}(2025)\citenamefont {Gil},
  \citenamefont {Oppel}, \citenamefont {Kottmann},\ and\ \citenamefont
  {Gonz{\'a}lez}}]{gil2025sharc}%
  \BibitemOpen
  \bibfield  {author} {\bibinfo {author} {\bibfnamefont {E.~S.}\ \bibnamefont
  {Gil}}, \bibinfo {author} {\bibfnamefont {M.}~\bibnamefont {Oppel}}, \bibinfo
  {author} {\bibfnamefont {J.~S.}\ \bibnamefont {Kottmann}},\ and\ \bibinfo
  {author} {\bibfnamefont {L.}~\bibnamefont {Gonz{\'a}lez}},\ }\bibfield
  {title} {\bibinfo {title} {Sharc meets tequila: mixed quantum-classical
  dynamics on a quantum computer using a hybrid quantum-classical algorithm},\
  }\href@noop {} {\bibfield  {journal} {\bibinfo  {journal} {Chemical Science}\
  }\textbf {\bibinfo {volume} {16}},\ \bibinfo {pages} {596} (\bibinfo {year}
  {2025})}\BibitemShut {NoStop}%
\bibitem [{\citenamefont {Santos}\ and\ \citenamefont
  {Kottmann}()}]{santosHybrid2024}%
  \BibitemOpen
  \bibfield  {author} {\bibinfo {author} {\bibfnamefont {F.~J. D.~A.}\
  \bibnamefont {Santos}}\ and\ \bibinfo {author} {\bibfnamefont {J.~S.}\
  \bibnamefont {Kottmann}},\ }\bibfield  {title} {\bibinfo {title} {A {{Hybrid
  Qubit Encoding}}: {{Splitting Fock Space}} into {{Fermionic}} and {{Bosonic
  Subspaces}}}\ }\href {https://doi.org/10.48550/arXiv.2411.14096}
  {10.48550/arXiv.2411.14096},\ \Eprint {https://arxiv.org/abs/2411.14096}
  {2411.14096} \BibitemShut {NoStop}%
\bibitem [{\citenamefont {Schleich}\ \emph {et~al.}(2023)\citenamefont
  {Schleich}, \citenamefont {Boen}, \citenamefont {Cincio}, \citenamefont
  {Anand}, \citenamefont {Kottmann}, \citenamefont {Tretiak}, \citenamefont
  {Dub},\ and\ \citenamefont {{Aspuru-Guzik}}}]{schleich2023partitioning}%
  \BibitemOpen
  \bibfield  {author} {\bibinfo {author} {\bibfnamefont {P.}~\bibnamefont
  {Schleich}}, \bibinfo {author} {\bibfnamefont {J.}~\bibnamefont {Boen}},
  \bibinfo {author} {\bibfnamefont {L.}~\bibnamefont {Cincio}}, \bibinfo
  {author} {\bibfnamefont {A.}~\bibnamefont {Anand}}, \bibinfo {author}
  {\bibfnamefont {J.~S.}\ \bibnamefont {Kottmann}}, \bibinfo {author}
  {\bibfnamefont {S.}~\bibnamefont {Tretiak}}, \bibinfo {author} {\bibfnamefont
  {P.~A.}\ \bibnamefont {Dub}},\ and\ \bibinfo {author} {\bibfnamefont
  {A.}~\bibnamefont {{Aspuru-Guzik}}},\ }\bibfield  {title} {\bibinfo {title}
  {Partitioning quantum chemistry simulations with clifford circuits},\ }\href
  {https://doi.org/10.1021/acs.jctc.3c00335} {\bibfield  {journal} {\bibinfo
  {journal} {Journal of Chemical Theory and Computation}\ }\textbf {\bibinfo
  {volume} {19}},\ \bibinfo {pages} {4952} (\bibinfo {year} {2023})},\ \Eprint
  {https://arxiv.org/abs/https://doi.org/10.1021/acs.jctc.3c00335}
  {https://doi.org/10.1021/acs.jctc.3c00335} \BibitemShut {NoStop}%
\bibitem [{\citenamefont {Kottmann}\ and\ \citenamefont
  {Scala}(2024)}]{kottmann2024quantum}%
  \BibitemOpen
  \bibfield  {author} {\bibinfo {author} {\bibfnamefont {J.~S.}\ \bibnamefont
  {Kottmann}}\ and\ \bibinfo {author} {\bibfnamefont {F.}~\bibnamefont
  {Scala}},\ }\bibfield  {title} {\bibinfo {title} {Quantum algorithmic
  approach to multiconfigurational valence bond theory: Insights from
  interpretable circuit design},\ }\href@noop {} {\bibfield  {journal}
  {\bibinfo  {journal} {Journal of Chemical Theory and Computation}\ }\textbf
  {\bibinfo {volume} {20}},\ \bibinfo {pages} {3514} (\bibinfo {year}
  {2024})}\BibitemShut {NoStop}%
\bibitem [{\citenamefont {McClean}\ \emph {et~al.}(2018)\citenamefont
  {McClean}, \citenamefont {Boixo}, \citenamefont {Smelyanskiy}, \citenamefont
  {Babbush},\ and\ \citenamefont {Neven}}]{mcclean2018barren}%
  \BibitemOpen
  \bibfield  {author} {\bibinfo {author} {\bibfnamefont {J.~R.}\ \bibnamefont
  {McClean}}, \bibinfo {author} {\bibfnamefont {S.}~\bibnamefont {Boixo}},
  \bibinfo {author} {\bibfnamefont {V.~N.}\ \bibnamefont {Smelyanskiy}},
  \bibinfo {author} {\bibfnamefont {R.}~\bibnamefont {Babbush}},\ and\ \bibinfo
  {author} {\bibfnamefont {H.}~\bibnamefont {Neven}},\ }\bibfield  {title}
  {\bibinfo {title} {Barren plateaus in quantum neural network training
  landscapes},\ }\href {https://doi.org/10.1038/s41467-018-07090-4} {\bibfield
  {journal} {\bibinfo  {journal} {Nature Communications}\ }\textbf {\bibinfo
  {volume} {9}},\ \bibinfo {pages} {1} (\bibinfo {year} {2018})}\BibitemShut
  {NoStop}%
\bibitem [{\citenamefont {Bittel}\ and\ \citenamefont
  {Kliesch}(2021)}]{bittel2021training}%
  \BibitemOpen
  \bibfield  {author} {\bibinfo {author} {\bibfnamefont {L.}~\bibnamefont
  {Bittel}}\ and\ \bibinfo {author} {\bibfnamefont {M.}~\bibnamefont
  {Kliesch}},\ }\bibfield  {title} {\bibinfo {title} {Training variational
  quantum algorithms is {{NP-hard}}\textendash even for logarithmically many
  qubits and free fermionic systems},\ }\href
  {https://arxiv.org/abs/2101.07267} {\bibfield  {journal} {\bibinfo  {journal}
  {arxiv preprint arxiv:2101.07267}\ } (\bibinfo {year} {2021})},\ \Eprint
  {https://arxiv.org/abs/2101.07267} {arxiv:2101.07267} \BibitemShut {NoStop}%
\bibitem [{\citenamefont {Zhang}\ and\ \citenamefont
  {Yin}(2020)}]{zhang2020collective}%
  \BibitemOpen
  \bibfield  {author} {\bibinfo {author} {\bibfnamefont {D.-B.}\ \bibnamefont
  {Zhang}}\ and\ \bibinfo {author} {\bibfnamefont {T.}~\bibnamefont {Yin}},\
  }\bibfield  {title} {\bibinfo {title} {Collective optimization for
  variational quantum eigensolvers},\ }\href
  {https://doi.org/10.1103/PhysRevA.101.032311} {\bibfield  {journal} {\bibinfo
   {journal} {Phys. Rev. A}\ }\textbf {\bibinfo {volume} {101}},\ \bibinfo
  {pages} {032311} (\bibinfo {year} {2020})}\BibitemShut {NoStop}%
\bibitem [{\citenamefont {Cervera-Lierta}\ \emph {et~al.}(2021)\citenamefont
  {Cervera-Lierta}, \citenamefont {Kottmann},\ and\ \citenamefont
  {Aspuru-Guzik}}]{MetaVQE}%
  \BibitemOpen
  \bibfield  {author} {\bibinfo {author} {\bibfnamefont {A.}~\bibnamefont
  {Cervera-Lierta}}, \bibinfo {author} {\bibfnamefont {J.~S.}\ \bibnamefont
  {Kottmann}},\ and\ \bibinfo {author} {\bibfnamefont {A.}~\bibnamefont
  {Aspuru-Guzik}},\ }\bibfield  {title} {\bibinfo {title} {Meta-variational
  quantum eigensolver: Learning energy profiles of parameterized hamiltonians
  for quantum simulation},\ }\bibfield  {journal} {\bibinfo  {journal} {PRX
  Quantum}\ }\textbf {\bibinfo {volume} {2}},\ \href
  {https://doi.org/10.1103/prxquantum.2.020329} {10.1103/prxquantum.2.020329}
  (\bibinfo {year} {2021})\BibitemShut {NoStop}%
\bibitem [{\citenamefont {Ceroni}\ \emph {et~al.}(2023)\citenamefont {Ceroni},
  \citenamefont {Stetina}, \citenamefont {Kieferova}, \citenamefont {Marrero},
  \citenamefont {Arrazola},\ and\ \citenamefont
  {Wiebe}}]{related-ceroni2023generatingapproximategroundstates}%
  \BibitemOpen
  \bibfield  {author} {\bibinfo {author} {\bibfnamefont {J.}~\bibnamefont
  {Ceroni}}, \bibinfo {author} {\bibfnamefont {T.~F.}\ \bibnamefont {Stetina}},
  \bibinfo {author} {\bibfnamefont {M.}~\bibnamefont {Kieferova}}, \bibinfo
  {author} {\bibfnamefont {C.~O.}\ \bibnamefont {Marrero}}, \bibinfo {author}
  {\bibfnamefont {J.~M.}\ \bibnamefont {Arrazola}},\ and\ \bibinfo {author}
  {\bibfnamefont {N.}~\bibnamefont {Wiebe}},\ }\href
  {https://arxiv.org/abs/2210.05489} {\bibinfo {title} {Generating approximate
  ground states of molecules using quantum machine learning}} (\bibinfo {year}
  {2023}),\ \Eprint {https://arxiv.org/abs/2210.05489} {arXiv:2210.05489
  [quant-ph]} \BibitemShut {NoStop}%
\bibitem [{\citenamefont {Truger}\ \emph {et~al.}(2024)\citenamefont {Truger},
  \citenamefont {Barzen}, \citenamefont {Leymann},\ and\ \citenamefont
  {Obst}}]{related-warm-starting-vqe}%
  \BibitemOpen
  \bibfield  {author} {\bibinfo {author} {\bibfnamefont {F.}~\bibnamefont
  {Truger}}, \bibinfo {author} {\bibfnamefont {J.}~\bibnamefont {Barzen}},
  \bibinfo {author} {\bibfnamefont {F.}~\bibnamefont {Leymann}},\ and\ \bibinfo
  {author} {\bibfnamefont {J.}~\bibnamefont {Obst}},\ }\href
  {https://arxiv.org/abs/2402.17378} {\bibinfo {title} {Warm-starting the vqe
  with approximate complex amplitude encoding}} (\bibinfo {year} {2024}),\
  \Eprint {https://arxiv.org/abs/2402.17378} {arXiv:2402.17378 [quant-ph]}
  \BibitemShut {NoStop}%
\bibitem [{\citenamefont {Nakaji}\ \emph {et~al.}(2024)\citenamefont {Nakaji},
  \citenamefont {Kristensen}, \citenamefont {Campos-Gonzalez-Angulo},
  \citenamefont {Vakili}, \citenamefont {Huang}, \citenamefont {Bagherimehrab},
  \citenamefont {Gorgulla}, \citenamefont {Wong}, \citenamefont {McCaskey},
  \citenamefont {Kim}, \citenamefont {Nguyen}, \citenamefont {Rao},\ and\
  \citenamefont {Aspuru-Guzik}}]{related-generative-quantum-eigensolver}%
  \BibitemOpen
  \bibfield  {author} {\bibinfo {author} {\bibfnamefont {K.}~\bibnamefont
  {Nakaji}}, \bibinfo {author} {\bibfnamefont {L.~B.}\ \bibnamefont
  {Kristensen}}, \bibinfo {author} {\bibfnamefont {J.~A.}\ \bibnamefont
  {Campos-Gonzalez-Angulo}}, \bibinfo {author} {\bibfnamefont {M.~G.}\
  \bibnamefont {Vakili}}, \bibinfo {author} {\bibfnamefont {H.}~\bibnamefont
  {Huang}}, \bibinfo {author} {\bibfnamefont {M.}~\bibnamefont
  {Bagherimehrab}}, \bibinfo {author} {\bibfnamefont {C.}~\bibnamefont
  {Gorgulla}}, \bibinfo {author} {\bibfnamefont {F.}~\bibnamefont {Wong}},
  \bibinfo {author} {\bibfnamefont {A.}~\bibnamefont {McCaskey}}, \bibinfo
  {author} {\bibfnamefont {J.-S.}\ \bibnamefont {Kim}}, \bibinfo {author}
  {\bibfnamefont {T.}~\bibnamefont {Nguyen}}, \bibinfo {author} {\bibfnamefont
  {P.}~\bibnamefont {Rao}},\ and\ \bibinfo {author} {\bibfnamefont
  {A.}~\bibnamefont {Aspuru-Guzik}},\ }\href {https://arxiv.org/abs/2401.09253}
  {\bibinfo {title} {The generative quantum eigensolver (gqe) and its
  application for ground state search}} (\bibinfo {year} {2024}),\ \Eprint
  {https://arxiv.org/abs/2401.09253} {arXiv:2401.09253 [quant-ph]} \BibitemShut
  {NoStop}%
\bibitem [{\citenamefont {Bensoussan}\ \emph {et~al.}(2025)\citenamefont
  {Bensoussan}, \citenamefont {Chachkarova}, \citenamefont {Even-Mendoza},
  \citenamefont {Fortz},\ and\ \citenamefont
  {Lenihan}}]{bensoussan2025accelerq}%
  \BibitemOpen
  \bibfield  {author} {\bibinfo {author} {\bibfnamefont {A.}~\bibnamefont
  {Bensoussan}}, \bibinfo {author} {\bibfnamefont {E.}~\bibnamefont
  {Chachkarova}}, \bibinfo {author} {\bibfnamefont {K.}~\bibnamefont
  {Even-Mendoza}}, \bibinfo {author} {\bibfnamefont {S.}~\bibnamefont
  {Fortz}},\ and\ \bibinfo {author} {\bibfnamefont {C.}~\bibnamefont
  {Lenihan}},\ }\bibfield  {title} {\bibinfo {title} {Accelerq: Accelerating
  quantum eigensolvers with machine learning on quantum simulators},\ }\href
  {https://doi.org/10.1145/3763132} {\bibfield  {journal} {\bibinfo  {journal}
  {Proc. ACM Program. Lang.}\ }\textbf {\bibinfo {volume} {9}},\ \bibinfo
  {pages} {2279} (\bibinfo {year} {2025})}\BibitemShut {NoStop}%
\bibitem [{\citenamefont {Veli{\v{c}}kovi{\'c}}\ \emph
  {et~al.}(2017)\citenamefont {Veli{\v{c}}kovi{\'c}}, \citenamefont {Cucurull},
  \citenamefont {Casanova}, \citenamefont {Romero}, \citenamefont {Lio},\ and\
  \citenamefont {Bengio}}]{gat}%
  \BibitemOpen
  \bibfield  {author} {\bibinfo {author} {\bibfnamefont {P.}~\bibnamefont
  {Veli{\v{c}}kovi{\'c}}}, \bibinfo {author} {\bibfnamefont {G.}~\bibnamefont
  {Cucurull}}, \bibinfo {author} {\bibfnamefont {A.}~\bibnamefont {Casanova}},
  \bibinfo {author} {\bibfnamefont {A.}~\bibnamefont {Romero}}, \bibinfo
  {author} {\bibfnamefont {P.}~\bibnamefont {Lio}},\ and\ \bibinfo {author}
  {\bibfnamefont {Y.}~\bibnamefont {Bengio}},\ }\bibfield  {title} {\bibinfo
  {title} {Graph attention networks},\ }\href@noop {} {\bibfield  {journal}
  {\bibinfo  {journal} {arXiv preprint arXiv:1710.10903}\ } (\bibinfo {year}
  {2017})}\BibitemShut {NoStop}%
\bibitem [{\citenamefont {Gilmer}\ \emph {et~al.}(2017)\citenamefont {Gilmer},
  \citenamefont {Schoenholz}, \citenamefont {Riley}, \citenamefont {Vinyals},\
  and\ \citenamefont {Dahl}}]{related-neural_message_passing}%
  \BibitemOpen
  \bibfield  {author} {\bibinfo {author} {\bibfnamefont {J.}~\bibnamefont
  {Gilmer}}, \bibinfo {author} {\bibfnamefont {S.~S.}\ \bibnamefont
  {Schoenholz}}, \bibinfo {author} {\bibfnamefont {P.~F.}\ \bibnamefont
  {Riley}}, \bibinfo {author} {\bibfnamefont {O.}~\bibnamefont {Vinyals}},\
  and\ \bibinfo {author} {\bibfnamefont {G.~E.}\ \bibnamefont {Dahl}},\ }\href
  {https://arxiv.org/abs/1704.01212} {\bibinfo {title} {Neural message passing
  for quantum chemistry}} (\bibinfo {year} {2017}),\ \Eprint
  {https://arxiv.org/abs/1704.01212} {arXiv:1704.01212 [cs.LG]} \BibitemShut
  {NoStop}%
\bibitem [{\citenamefont {Schütt}\ \emph {et~al.}(2017)\citenamefont
  {Schütt}, \citenamefont {Kindermans}, \citenamefont {Sauceda}, \citenamefont
  {Chmiela}, \citenamefont {Tkatchenko},\ and\ \citenamefont
  {Müller}}]{related-schnet}%
  \BibitemOpen
  \bibfield  {author} {\bibinfo {author} {\bibfnamefont {K.~T.}\ \bibnamefont
  {Schütt}}, \bibinfo {author} {\bibfnamefont {P.-J.}\ \bibnamefont
  {Kindermans}}, \bibinfo {author} {\bibfnamefont {H.~E.}\ \bibnamefont
  {Sauceda}}, \bibinfo {author} {\bibfnamefont {S.}~\bibnamefont {Chmiela}},
  \bibinfo {author} {\bibfnamefont {A.}~\bibnamefont {Tkatchenko}},\ and\
  \bibinfo {author} {\bibfnamefont {K.-R.}\ \bibnamefont {Müller}},\ }\href
  {https://arxiv.org/abs/1706.08566} {\bibinfo {title} {Schnet: A
  continuous-filter convolutional neural network for modeling quantum
  interactions}} (\bibinfo {year} {2017}),\ \Eprint
  {https://arxiv.org/abs/1706.08566} {arXiv:1706.08566 [stat.ML]} \BibitemShut
  {NoStop}%
\bibitem [{\citenamefont {von Glehn}\ \emph {et~al.}(2023)\citenamefont {von
  Glehn}, \citenamefont {Spencer},\ and\ \citenamefont
  {Pfau}}]{related-psiformer-attention}%
  \BibitemOpen
  \bibfield  {author} {\bibinfo {author} {\bibfnamefont {I.}~\bibnamefont {von
  Glehn}}, \bibinfo {author} {\bibfnamefont {J.~S.}\ \bibnamefont {Spencer}},\
  and\ \bibinfo {author} {\bibfnamefont {D.}~\bibnamefont {Pfau}},\ }\href
  {https://arxiv.org/abs/2211.13672} {\bibinfo {title} {A self-attention ansatz
  for ab-initio quantum chemistry}} (\bibinfo {year} {2023}),\ \Eprint
  {https://arxiv.org/abs/2211.13672} {arXiv:2211.13672 [physics.chem-ph]}
  \BibitemShut {NoStop}%
\bibitem [{\citenamefont {Sch{\"u}tt}\ \emph {et~al.}(2017)\citenamefont
  {Sch{\"u}tt}, \citenamefont {Kindermans}, \citenamefont {Sauceda~Felix},
  \citenamefont {Chmiela}, \citenamefont {Tkatchenko},\ and\ \citenamefont
  {M{\"u}ller}}]{schnet}%
  \BibitemOpen
  \bibfield  {author} {\bibinfo {author} {\bibfnamefont {K.}~\bibnamefont
  {Sch{\"u}tt}}, \bibinfo {author} {\bibfnamefont {P.-J.}\ \bibnamefont
  {Kindermans}}, \bibinfo {author} {\bibfnamefont {H.~E.}\ \bibnamefont
  {Sauceda~Felix}}, \bibinfo {author} {\bibfnamefont {S.}~\bibnamefont
  {Chmiela}}, \bibinfo {author} {\bibfnamefont {A.}~\bibnamefont
  {Tkatchenko}},\ and\ \bibinfo {author} {\bibfnamefont {K.-R.}\ \bibnamefont
  {M{\"u}ller}},\ }\bibfield  {title} {\bibinfo {title} {Schnet: A
  continuous-filter convolutional neural network for modeling quantum
  interactions},\ }\href@noop {} {\bibfield  {journal} {\bibinfo  {journal}
  {Advances in neural information processing systems}\ }\textbf {\bibinfo
  {volume} {30}} (\bibinfo {year} {2017})}\BibitemShut {NoStop}%
\bibitem [{\citenamefont {Stein}(2024)}]{steinNylser2025}%
  \BibitemOpen
  \bibfield  {author} {\bibinfo {author} {\bibfnamefont {K.}~\bibnamefont
  {Stein}},\ }\href {https://github.com/nylser/quanti-gin} {\bibinfo {title}
  {Nylser/quanti-gin}} (\bibinfo {year} {2024})\BibitemShut {NoStop}%
\bibitem [{\citenamefont {Kottmann}\ \emph
  {et~al.}(2021{\natexlab{a}})\citenamefont {Kottmann}, \citenamefont
  {Alperin-Lea}, \citenamefont {Tamayo-Mendoza}, \citenamefont
  {Cervera-Lierta}, \citenamefont {Lavigne}, \citenamefont {Yen}, \citenamefont
  {Verteletskyi}, \citenamefont {Schleich}, \citenamefont {Anand},
  \citenamefont {Degroote}, \citenamefont {Chaney}, \citenamefont {Kesibi},
  \citenamefont {Curnow}, \citenamefont {Solo}, \citenamefont
  {Tsilimigkounakis}, \citenamefont {Zendejas-Morales}, \citenamefont
  {Izmaylov},\ and\ \citenamefont {Aspuru-Guzik}}]{tequila}%
  \BibitemOpen
  \bibfield  {author} {\bibinfo {author} {\bibfnamefont {J.~S.}\ \bibnamefont
  {Kottmann}}, \bibinfo {author} {\bibfnamefont {S.}~\bibnamefont
  {Alperin-Lea}}, \bibinfo {author} {\bibfnamefont {T.}~\bibnamefont
  {Tamayo-Mendoza}}, \bibinfo {author} {\bibfnamefont {A.}~\bibnamefont
  {Cervera-Lierta}}, \bibinfo {author} {\bibfnamefont {C.}~\bibnamefont
  {Lavigne}}, \bibinfo {author} {\bibfnamefont {T.-C.}\ \bibnamefont {Yen}},
  \bibinfo {author} {\bibfnamefont {V.}~\bibnamefont {Verteletskyi}}, \bibinfo
  {author} {\bibfnamefont {P.}~\bibnamefont {Schleich}}, \bibinfo {author}
  {\bibfnamefont {A.}~\bibnamefont {Anand}}, \bibinfo {author} {\bibfnamefont
  {M.}~\bibnamefont {Degroote}}, \bibinfo {author} {\bibfnamefont
  {S.}~\bibnamefont {Chaney}}, \bibinfo {author} {\bibfnamefont
  {M.}~\bibnamefont {Kesibi}}, \bibinfo {author} {\bibfnamefont {N.~G.}\
  \bibnamefont {Curnow}}, \bibinfo {author} {\bibfnamefont {B.}~\bibnamefont
  {Solo}}, \bibinfo {author} {\bibfnamefont {G.}~\bibnamefont
  {Tsilimigkounakis}}, \bibinfo {author} {\bibfnamefont {C.}~\bibnamefont
  {Zendejas-Morales}}, \bibinfo {author} {\bibfnamefont {A.~F.}\ \bibnamefont
  {Izmaylov}},\ and\ \bibinfo {author} {\bibfnamefont {A.}~\bibnamefont
  {Aspuru-Guzik}},\ }\bibfield  {title} {\bibinfo {title} {{{TEQUILA}}: A
  platform for rapid development of quantum algorithms},\ }\href
  {https://doi.org/10.1088/2058-9565/abe567} {\bibfield  {journal} {\bibinfo
  {journal} {Quantum Science and Technology}\ }\textbf {\bibinfo {volume}
  {6}},\ \bibinfo {pages} {024009} (\bibinfo {year}
  {2021}{\natexlab{a}})}\BibitemShut {NoStop}%
\bibitem [{\citenamefont {Kottmann}\ \emph
  {et~al.}(2021{\natexlab{b}})\citenamefont {Kottmann}, \citenamefont
  {Schleich}, \citenamefont {{Tamayo-Mendoza}},\ and\ \citenamefont
  {{Aspuru-Guzik}}}]{kottmann2020reducing}%
  \BibitemOpen
  \bibfield  {author} {\bibinfo {author} {\bibfnamefont {J.~S.}\ \bibnamefont
  {Kottmann}}, \bibinfo {author} {\bibfnamefont {P.}~\bibnamefont {Schleich}},
  \bibinfo {author} {\bibfnamefont {T.}~\bibnamefont {{Tamayo-Mendoza}}},\ and\
  \bibinfo {author} {\bibfnamefont {A.}~\bibnamefont {{Aspuru-Guzik}}},\
  }\bibfield  {title} {\bibinfo {title} {Reducing qubit requirements while
  maintaining numerical precision for the variational quantum eigensolver:
  {{A}} basis-set-free approach},\ }\href
  {https://doi.org/10.1021/acs.jpclett.0c03410} {\bibfield  {journal} {\bibinfo
   {journal} {Journal of Physical Chemistry Letters}\ }\textbf {\bibinfo
  {volume} {12}},\ \bibinfo {pages} {663} (\bibinfo {year}
  {2021}{\natexlab{b}})}\BibitemShut {NoStop}%
\bibitem [{\citenamefont {Bincoletto}\ and\ \citenamefont
  {Kottmann}(2025)}]{BincolettoKottmann2025}%
  \BibitemOpen
  \bibfield  {author} {\bibinfo {author} {\bibfnamefont {D.}~\bibnamefont
  {Bincoletto}}\ and\ \bibinfo {author} {\bibfnamefont {J.~S.}\ \bibnamefont
  {Kottmann}},\ }\bibfield  {title} {\bibinfo {title} {State specific
  measurement protocols for the variational quantum eigensolver}\ }\href
  {https://doi.org/10.48550/arXiv.2504.03019} {10.48550/arXiv.2504.03019}
  (\bibinfo {year} {2025}),\ \bibinfo {note} {preprint},\ \Eprint
  {https://arxiv.org/abs/arXiv:2504.03019} {arXiv:arXiv:2504.03019 [quant-ph]}
  \BibitemShut {NoStop}%
\bibitem [{\citenamefont {Ansel}\ \emph {et~al.}(2024)\citenamefont {Ansel},
  \citenamefont {Yang}, \citenamefont {He}, \citenamefont {Gimelshein},
  \citenamefont {Jain}, \citenamefont {Voznesensky}, \citenamefont {Bao},
  \citenamefont {Bell}, \citenamefont {Berard}, \citenamefont {Burovski},
  \citenamefont {Chauhan}, \citenamefont {Chourdia}, \citenamefont {Constable},
  \citenamefont {Desmaison}, \citenamefont {DeVito}, \citenamefont {Ellison},
  \citenamefont {Feng}, \citenamefont {Gong}, \citenamefont {Gschwind},
  \citenamefont {Hirsh}, \citenamefont {Huang}, \citenamefont {Kalambarkar},
  \citenamefont {Kirsch}, \citenamefont {Lazos}, \citenamefont {Lezcano},
  \citenamefont {Liang}, \citenamefont {Liang}, \citenamefont {Lu},
  \citenamefont {Luk}, \citenamefont {Maher}, \citenamefont {Pan},
  \citenamefont {Puhrsch}, \citenamefont {Reso}, \citenamefont {Saroufim},
  \citenamefont {Siraichi}, \citenamefont {Suk}, \citenamefont {Zhang},
  \citenamefont {Suo}, \citenamefont {Tillet}, \citenamefont {Zhao},
  \citenamefont {Wang}, \citenamefont {Zhou}, \citenamefont {Zou},
  \citenamefont {Wang}, \citenamefont {Mathews}, \citenamefont {Wen},
  \citenamefont {Chanan}, \citenamefont {Wu},\ and\ \citenamefont
  {Chintala}}]{pytorch}%
  \BibitemOpen
  \bibfield  {author} {\bibinfo {author} {\bibfnamefont {J.}~\bibnamefont
  {Ansel}}, \bibinfo {author} {\bibfnamefont {E.}~\bibnamefont {Yang}},
  \bibinfo {author} {\bibfnamefont {H.}~\bibnamefont {He}}, \bibinfo {author}
  {\bibfnamefont {N.}~\bibnamefont {Gimelshein}}, \bibinfo {author}
  {\bibfnamefont {A.}~\bibnamefont {Jain}}, \bibinfo {author} {\bibfnamefont
  {M.}~\bibnamefont {Voznesensky}}, \bibinfo {author} {\bibfnamefont
  {B.}~\bibnamefont {Bao}}, \bibinfo {author} {\bibfnamefont {P.}~\bibnamefont
  {Bell}}, \bibinfo {author} {\bibfnamefont {D.}~\bibnamefont {Berard}},
  \bibinfo {author} {\bibfnamefont {E.}~\bibnamefont {Burovski}}, \bibinfo
  {author} {\bibfnamefont {G.}~\bibnamefont {Chauhan}}, \bibinfo {author}
  {\bibfnamefont {A.}~\bibnamefont {Chourdia}}, \bibinfo {author}
  {\bibfnamefont {W.}~\bibnamefont {Constable}}, \bibinfo {author}
  {\bibfnamefont {A.}~\bibnamefont {Desmaison}}, \bibinfo {author}
  {\bibfnamefont {Z.}~\bibnamefont {DeVito}}, \bibinfo {author} {\bibfnamefont
  {E.}~\bibnamefont {Ellison}}, \bibinfo {author} {\bibfnamefont
  {W.}~\bibnamefont {Feng}}, \bibinfo {author} {\bibfnamefont {J.}~\bibnamefont
  {Gong}}, \bibinfo {author} {\bibfnamefont {M.}~\bibnamefont {Gschwind}},
  \bibinfo {author} {\bibfnamefont {B.}~\bibnamefont {Hirsh}}, \bibinfo
  {author} {\bibfnamefont {S.}~\bibnamefont {Huang}}, \bibinfo {author}
  {\bibfnamefont {K.}~\bibnamefont {Kalambarkar}}, \bibinfo {author}
  {\bibfnamefont {L.}~\bibnamefont {Kirsch}}, \bibinfo {author} {\bibfnamefont
  {M.}~\bibnamefont {Lazos}}, \bibinfo {author} {\bibfnamefont
  {M.}~\bibnamefont {Lezcano}}, \bibinfo {author} {\bibfnamefont
  {Y.}~\bibnamefont {Liang}}, \bibinfo {author} {\bibfnamefont
  {J.}~\bibnamefont {Liang}}, \bibinfo {author} {\bibfnamefont
  {Y.}~\bibnamefont {Lu}}, \bibinfo {author} {\bibfnamefont {C.~K.}\
  \bibnamefont {Luk}}, \bibinfo {author} {\bibfnamefont {B.}~\bibnamefont
  {Maher}}, \bibinfo {author} {\bibfnamefont {Y.}~\bibnamefont {Pan}}, \bibinfo
  {author} {\bibfnamefont {C.}~\bibnamefont {Puhrsch}}, \bibinfo {author}
  {\bibfnamefont {M.}~\bibnamefont {Reso}}, \bibinfo {author} {\bibfnamefont
  {M.}~\bibnamefont {Saroufim}}, \bibinfo {author} {\bibfnamefont {M.~Y.}\
  \bibnamefont {Siraichi}}, \bibinfo {author} {\bibfnamefont {H.}~\bibnamefont
  {Suk}}, \bibinfo {author} {\bibfnamefont {S.}~\bibnamefont {Zhang}}, \bibinfo
  {author} {\bibfnamefont {M.}~\bibnamefont {Suo}}, \bibinfo {author}
  {\bibfnamefont {P.}~\bibnamefont {Tillet}}, \bibinfo {author} {\bibfnamefont
  {X.}~\bibnamefont {Zhao}}, \bibinfo {author} {\bibfnamefont {E.}~\bibnamefont
  {Wang}}, \bibinfo {author} {\bibfnamefont {K.}~\bibnamefont {Zhou}}, \bibinfo
  {author} {\bibfnamefont {R.}~\bibnamefont {Zou}}, \bibinfo {author}
  {\bibfnamefont {X.}~\bibnamefont {Wang}}, \bibinfo {author} {\bibfnamefont
  {A.}~\bibnamefont {Mathews}}, \bibinfo {author} {\bibfnamefont
  {W.}~\bibnamefont {Wen}}, \bibinfo {author} {\bibfnamefont {G.}~\bibnamefont
  {Chanan}}, \bibinfo {author} {\bibfnamefont {P.}~\bibnamefont {Wu}},\ and\
  \bibinfo {author} {\bibfnamefont {S.}~\bibnamefont {Chintala}},\ }\bibfield
  {title} {\bibinfo {title} {Pytorch 2: Faster machine learning through dynamic
  python bytecode transformation and graph compilation},\ }in\ \href
  {https://doi.org/10.1145/3620665.3640366} {\emph {\bibinfo {booktitle}
  {Proceedings of the 29th ACM International Conference on Architectural
  Support for Programming Languages and Operating Systems, Volume 2}}},\
  \bibinfo {series and number} {ASPLOS '24}\ (\bibinfo  {publisher}
  {Association for Computing Machinery},\ \bibinfo {address} {New York, NY,
  USA},\ \bibinfo {year} {2024})\ p.\ \bibinfo {pages} {929–947}\BibitemShut
  {NoStop}%
\bibitem [{\citenamefont {Fey}\ and\ \citenamefont
  {Lenssen}(2019)}]{pytorch-geometric}%
  \BibitemOpen
  \bibfield  {author} {\bibinfo {author} {\bibfnamefont {M.}~\bibnamefont
  {Fey}}\ and\ \bibinfo {author} {\bibfnamefont {J.~E.}\ \bibnamefont
  {Lenssen}},\ }\href {https://arxiv.org/abs/1903.02428} {\bibinfo {title}
  {Fast graph representation learning with pytorch geometric}} (\bibinfo {year}
  {2019}),\ \Eprint {https://arxiv.org/abs/1903.02428} {arXiv:1903.02428
  [cs.LG]} \BibitemShut {NoStop}%
\bibitem [{\citenamefont {Vaswani}\ \emph {et~al.}(2017)\citenamefont
  {Vaswani}, \citenamefont {Shazeer}, \citenamefont {Parmar}, \citenamefont
  {Uszkoreit}, \citenamefont {Jones}, \citenamefont {Gomez}, \citenamefont
  {Kaiser},\ and\ \citenamefont
  {Polosukhin}}]{related-attention-is-all-you-need}%
  \BibitemOpen
  \bibfield  {author} {\bibinfo {author} {\bibfnamefont {A.}~\bibnamefont
  {Vaswani}}, \bibinfo {author} {\bibfnamefont {N.}~\bibnamefont {Shazeer}},
  \bibinfo {author} {\bibfnamefont {N.}~\bibnamefont {Parmar}}, \bibinfo
  {author} {\bibfnamefont {J.}~\bibnamefont {Uszkoreit}}, \bibinfo {author}
  {\bibfnamefont {L.}~\bibnamefont {Jones}}, \bibinfo {author} {\bibfnamefont
  {A.~N.}\ \bibnamefont {Gomez}}, \bibinfo {author} {\bibfnamefont
  {L.}~\bibnamefont {Kaiser}},\ and\ \bibinfo {author} {\bibfnamefont
  {I.}~\bibnamefont {Polosukhin}},\ }\bibfield  {title} {\bibinfo {title}
  {Attention is all you need},\ }\href {http://arxiv.org/abs/1706.03762}
  {\bibfield  {journal} {\bibinfo  {journal} {CoRR}\ }\textbf {\bibinfo
  {volume} {abs/1706.03762}} (\bibinfo {year} {2017})},\ \Eprint
  {https://arxiv.org/abs/1706.03762} {1706.03762} \BibitemShut {NoStop}%
\bibitem [{\citenamefont {Lee}\ \emph {et~al.}(2019)\citenamefont {Lee},
  \citenamefont {Rossi}, \citenamefont {Kim}, \citenamefont {Ahmed},\ and\
  \citenamefont {Koh}}]{survey-attention-models-graph}%
  \BibitemOpen
  \bibfield  {author} {\bibinfo {author} {\bibfnamefont {J.~B.}\ \bibnamefont
  {Lee}}, \bibinfo {author} {\bibfnamefont {R.~A.}\ \bibnamefont {Rossi}},
  \bibinfo {author} {\bibfnamefont {S.}~\bibnamefont {Kim}}, \bibinfo {author}
  {\bibfnamefont {N.~K.}\ \bibnamefont {Ahmed}},\ and\ \bibinfo {author}
  {\bibfnamefont {E.}~\bibnamefont {Koh}},\ }\bibfield  {title} {\bibinfo
  {title} {Attention models in graphs: A survey},\ }\bibfield  {journal}
  {\bibinfo  {journal} {ACM Trans. Knowl. Discov. Data}\ }\textbf {\bibinfo
  {volume} {13}},\ \href {https://doi.org/10.1145/3363574} {10.1145/3363574}
  (\bibinfo {year} {2019})\BibitemShut {NoStop}%
\bibitem [{\citenamefont {Veličković}\ \emph {et~al.}(2018)\citenamefont
  {Veličković}, \citenamefont {Cucurull}, \citenamefont {Casanova},
  \citenamefont {Romero}, \citenamefont {Liò},\ and\ \citenamefont
  {Bengio}}]{gatconv}%
  \BibitemOpen
  \bibfield  {author} {\bibinfo {author} {\bibfnamefont {P.}~\bibnamefont
  {Veličković}}, \bibinfo {author} {\bibfnamefont {G.}~\bibnamefont
  {Cucurull}}, \bibinfo {author} {\bibfnamefont {A.}~\bibnamefont {Casanova}},
  \bibinfo {author} {\bibfnamefont {A.}~\bibnamefont {Romero}}, \bibinfo
  {author} {\bibfnamefont {P.}~\bibnamefont {Liò}},\ and\ \bibinfo {author}
  {\bibfnamefont {Y.}~\bibnamefont {Bengio}},\ }\href
  {https://arxiv.org/abs/1710.10903} {\bibinfo {title} {Graph attention
  networks}} (\bibinfo {year} {2018}),\ \Eprint
  {https://arxiv.org/abs/1710.10903} {arXiv:1710.10903 [stat.ML]} \BibitemShut
  {NoStop}%
\bibitem [{\citenamefont {He}\ \emph {et~al.}(2016)\citenamefont {He},
  \citenamefont {Zhang}, \citenamefont {Ren},\ and\ \citenamefont
  {Sun}}]{resnet}%
  \BibitemOpen
  \bibfield  {author} {\bibinfo {author} {\bibfnamefont {K.}~\bibnamefont
  {He}}, \bibinfo {author} {\bibfnamefont {X.}~\bibnamefont {Zhang}}, \bibinfo
  {author} {\bibfnamefont {S.}~\bibnamefont {Ren}},\ and\ \bibinfo {author}
  {\bibfnamefont {J.}~\bibnamefont {Sun}},\ }\bibfield  {title} {\bibinfo
  {title} {Deep residual learning for image recognition},\ }in\ \href@noop {}
  {\emph {\bibinfo {booktitle} {Proceedings of the IEEE conference on computer
  vision and pattern recognition}}}\ (\bibinfo {year} {2016})\ pp.\ \bibinfo
  {pages} {770--778}\BibitemShut {NoStop}%
\bibitem [{\citenamefont {Anselmetti}\ \emph {et~al.}(2021)\citenamefont
  {Anselmetti}, \citenamefont {Wierichs}, \citenamefont {Gogolin},\ and\
  \citenamefont {Parrish}}]{anselmetti2021local}%
  \BibitemOpen
  \bibfield  {author} {\bibinfo {author} {\bibfnamefont {G.-L.~R.}\
  \bibnamefont {Anselmetti}}, \bibinfo {author} {\bibfnamefont
  {D.}~\bibnamefont {Wierichs}}, \bibinfo {author} {\bibfnamefont
  {C.}~\bibnamefont {Gogolin}},\ and\ \bibinfo {author} {\bibfnamefont {R.~M.}\
  \bibnamefont {Parrish}},\ }\bibfield  {title} {\bibinfo {title} {Local,
  expressive, quantum-number-preserving vqe ansätze for fermionic systems},\
  }\href {https://doi.org/10.1088/1367-2630/ac2cb3} {\bibfield  {journal}
  {\bibinfo  {journal} {New Journal of Physics}\ }\textbf {\bibinfo {volume}
  {23}},\ \bibinfo {pages} {113010} (\bibinfo {year} {2021})}\BibitemShut
  {NoStop}%
\bibitem [{\citenamefont {Cortes}\ \emph {et~al.}()\citenamefont {Cortes},
  \citenamefont {Rocca}, \citenamefont {Gonthier}, \citenamefont {Ollitrault},
  \citenamefont {Parrish}, \citenamefont {Anselmetti}, \citenamefont
  {Degroote}, \citenamefont {Moll}, \citenamefont {Santagati},\ and\
  \citenamefont {Streif}}]{cortesAssessing2024}%
  \BibitemOpen
  \bibfield  {author} {\bibinfo {author} {\bibfnamefont {C.~L.}\ \bibnamefont
  {Cortes}}, \bibinfo {author} {\bibfnamefont {D.}~\bibnamefont {Rocca}},
  \bibinfo {author} {\bibfnamefont {J.}~\bibnamefont {Gonthier}}, \bibinfo
  {author} {\bibfnamefont {P.~J.}\ \bibnamefont {Ollitrault}}, \bibinfo
  {author} {\bibfnamefont {R.~M.}\ \bibnamefont {Parrish}}, \bibinfo {author}
  {\bibfnamefont {G.-L.~R.}\ \bibnamefont {Anselmetti}}, \bibinfo {author}
  {\bibfnamefont {M.}~\bibnamefont {Degroote}}, \bibinfo {author}
  {\bibfnamefont {N.}~\bibnamefont {Moll}}, \bibinfo {author} {\bibfnamefont
  {R.}~\bibnamefont {Santagati}},\ and\ \bibinfo {author} {\bibfnamefont
  {M.}~\bibnamefont {Streif}},\ }\bibfield  {title} {\bibinfo {title}
  {Assessing the query complexity limits of quantum phase estimation using
  symmetry aware spectral bounds},\ }\href {http://arxiv.org/abs/2403.04737} {\
  }\Eprint {https://arxiv.org/abs/2403.04737} {2403.04737} \BibitemShut
  {NoStop}%
\end{thebibliography}%

\clearpage
\onecolumngrid
\begin{appendix}
\section{Variational Quantum Eigensolvers}
\label{c:background}
\label{c:VQE}

Given that the expectation value of an Hamiltonian can be expressed as a sum of expectation values of its components
\begin{equation}
  \expect{H} = \sum_{i\alpha} h^i_\alpha \expect{\sigma^i_\alpha} + \sum_{i j \alpha \beta} h^{ij}_{\alpha\beta} \expect{\sigma^i_\alpha \sigma^j_\beta} + \dots
\end{equation}
with $h \in \mathrm{R}$ being real numbers and $\sigma$ Pauli operators, the VQE routine can be summarized with the following steps:
\begin{enumerate}
  \item Prepare a quantum state $|\Psi(\vec{\theta})\rangle$ with a parametrized ansatz circuit on the quantum computer
  \item Measure the expectation values $\expect{H_i}$ and combine them classically to approximate $\expect{H}$.
  \item Update the parameters $\vec{\theta}$ in the state preparation, by minimizing $\expect{H}$ with an optimization algorithm, e.g., gradient descent.
\end{enumerate}

VQEs require lower quantum resources in terms of circuit depth and qubit coherence time than traditional quantum algorithms such as Quantum Phase Estimation. \cite{aspuru2005simulated, cortesAssessing2024} This is due to a hybrid quantum-classical structure. The relevant aspect of VQE is the decomposition and computation of the expectation value of an operator as the sum of individual or grouped expectation values. The trade-off of such algorithm is a significant overhead due to measurement repetitions and classical processing. Moreover, the measurement procedure can be affected by noise, which can impact its accuracy and precision.

\section{The Separable Pair Model}
\label{c:SPA}
In the following we will give a high-level description of the SPA -- a variational circuit design that has been proven effective with regards to convergence rate and consistency. We will limit ourselves to hydrogenic systems in minimal atomic bases and refer to the original works for general molecules.~\cite{kottmann2022optimized, kottmannMolecular2023, kottmann2024quantum}\\

Hydrogenic systems are a collection of hydrogen atoms in 3D (spatial) space. For $N$ hydrogen atoms the input data consists solely of $N$ 3-tuples representing the spatial coordinates. 
For each of the $N$ atoms a 3D function, called basis function, is assigned. This basis functions resemble the ground state of the isolated hydrogen atom and are used to define the electronic Hamiltonian. For simplicity we can imagine each of these orbitals as quantum state for an electron localized at one of the atoms.

In this work, we use the approximation which assumes that the total wavefunction is a product of individual electron pairs and comes in a highly optimized form with regards to gate-count and circuit-depth. In contrast to other VQE variants, SPAs can be automated and have shown to give consistent results. They are however not a isolated method but can be identified as sub-structures in various VQE approaches, such as improved graph-based designs~\cite{kottmannMolecular2023}, tiled-unitary products~\cite{burton2024accurate}, quantum number preserving fabrics~\cite{anselmetti2021local} or decomposed approaches constructed form linear combinations over different SPA circuits~\cite{kottmann2024quantum}. 

The energy expectation values of SPA circuits are not invariant with respect to orbital rotations -- meaning unitary rotations in the space of the basis functions that define the Hamiltonian can lead to better energies. This is not a surprise, as changing the basis functions also changes the interpretation which orbitals are assigned to the individual electron pairs. Note that these basis rotations do not change the spectrum of the Hamiltonian itself, they merely improve the quality of the separable pair approximation. As they can be optimized classically, we do not explicitly include them into the circuit within this work.

\begin{algorithm}[H]
\caption{Linear Coordinates Generation Along the $z$-Axis}
\label{alg:linear-coordinates}
\KwIn{
    Number of atoms $n \in \mathbb{N}$, \\
    iteration index $k \in \{0, \dots, T-1\}$ of $T \in \mathbb{N}$, \\
    boundaries $d_{\min}, \: d_{\max} \in \mathbb{R}^+$
}
\KwOut{
    Coordinate matrix $C \in \mathbb{R}^{n \times 3}$
}
\BlankLine
$d_{\mathrm{step}} \gets d_{\min} + (d_{\max} - d_{\min}) \cdot \frac{k}{T-1}$\\

$z_i \gets i \cdot d_{\mathrm{step}}, \quad i = 0, 1, \dots, n-1$

Initialize $C \gets \mathbf{0}_{n \times 3}$\; 

Set third column of $C$ to $(z_0, z_1, \dots, z_{n-1})$\; 

\Return $C$\;
\end{algorithm}

\begin{algorithm}[H]
\caption{Ring Coordinates Generation in the $xy$-Plane}
\label{alg:ring-coordinates}
\KwIn{
    Number of atoms $n \in \mathbb{N}$, \\
    iteration index $k \in \{0, \dots, T-1\}$ of $T \in \mathbb{N}$, \\
    boundaries $d_{\min}, \: d_{\max} \in \mathbb{R}^+$
}
\KwOut{
    Coordinate matrix $C \in \mathbb{R}^{n \times 3}$
}
\BlankLine

$d_{\mathrm{step}} \gets d_{\min} + (d_{\max} - d_{\min}) \cdot \frac{k}{T-1}$

$r \gets \frac{d_{\mathrm{step}}}{2 \,\sin\!\left(\tfrac{\pi}{n}\right)}$

Initialize $C \gets \mathbf{0}_{n \times 3}$\;

\For{$i \gets 0$ \KwTo $n-1$}{
    $\theta \gets \tfrac{2\pi i}{n}$\;
    
    $x \gets r \cdot \cos(\theta)$\;
    
    $y \gets r \cdot \sin(\theta)$\;
    
    Set $C[i,0] \gets x$, $C[i,1] \gets y$, $C[i,2] \gets 0$\; 
}
\Return $C$\;
\end{algorithm}

\clearpage
\begin{minipage}{\linewidth}
\label{lst:h4-example}
\begin{lstlisting}[language=Python, caption=Tequila H4 example, firstline=7, lastline=43, showstringspaces=false, basicstyle=\ttfamily\footnotesize]
def calculate_optimization():
    geometry = """
H 1.0 1.0 1.0
H 2.0 1.0 1.0
H 4.0 2.0 2.0
H 5.0 2.0 2.0"""

    edges = [[0, 1], [2, 3]]
    mol = tq.Molecule(geometry=geometry, basis_set="sto-3g")

    U = mol.make_ansatz("HCB-SPA", edges=edges)
    export_to(U, filename="h4_circuit.pdf")

    # construct initial guess
    initial_guess = np.eye(4)
    for edge in edges:
        initial_guess[edge[0]][edge[0]] = 1.0
        initial_guess[edge[0]][edge[1]] = 1.0
        initial_guess[edge[1]][edge[0]] = -1.0
        initial_guess[edge[1]][edge[1]] = 1.0

    opt_orbitals = tq.chemistry.optimize_orbitals(
        mol.use_native_orbitals(),
        U,
        initial_guess=initial_guess,
        use_hcb=True,
        silent=True,
    )

    print("\n\n\n====optimized orbitals====\n\n\n", opt_orbitals.mo_coeff)

    mol = opt_orbitals.molecule
    H = mol.make_hardcore_boson_hamiltonian()

    print("\n\n\n====running optimization====\n\n\n")
    E = tq.ExpectationValue(H=H, U=U)
    result = tq.minimize(E, silent=True)
\end{lstlisting}
\end{minipage}

\end{appendix}
\end{document}